\documentclass[aps,prb,twocolumn,showpacs,amsmath,floatfix]{revtex4-2}

\usepackage{physics}
\usepackage{tikz}
\usepackage{siunitx}
\usetikzlibrary{arrows.meta}
\usepackage{zref-xr}
\usepackage{graphicx}
\usepackage{booktabs}
\usepackage{epstopdf}
\usepackage{placeins}
\usepackage{marvosym}
\usepackage{amsmath}
\usepackage{tensor}
\usepackage[utf8]{inputenc}
\usepackage{pgfplots}
\pgfplotsset{compat=newest}
\usepackage{amssymb}
\usepackage{amsfonts}
\usepackage{multirow}
\usepackage{bm}
\usepackage{dcolumn}
\usepackage[caption=false]{subfig}
\usepackage{upgreek}
\usepackage[euler]{textgreek}
\usepackage{braket}
\usepackage[table]{xcolor}
\usepackage{hyperref}
\hypersetup{
	colorlinks=true,
	citecolor=blue,
	linkcolor=purple,
	filecolor=magenta,
	urlcolor=black,
}
\usepackage{enumitem}
\usepackage{etaremune}
\usepackage{longtable}
\usepackage{array}
\usepackage{titlecaps}

\let\sidesubfloat\subfloat

\newcommand{\etal}{\textit{et al.\ }}

\begin{document}
	
	
	\title{Electronic and optical properties of arsenic monolayer: from planar honeycomb to the puckered phase}
	
	\date{\today}
	
	\author{Niloufar Dadkhah}
	\email{niloufar.dadkhah@case.edu}
	\author{Walter R. L. Lambrecht}
	\email{walter.lambrecht@case.edu}
	\affiliation{Department of Physics, Case Western Reserve University, 10900 Euclid Avenue, Cleveland, Ohio 44106-7079, USA}
	
	\begin{abstract}
		Group-V monolayer materials exhibit intriguing electronic and optical properties, influenced by their unique crystal symmetries and structural phases. In this work, we study arsenic monolayers, investigating their electronic and optical  properties across different phases, including planar, and puckered forms, using density functional theory (DFT) and quasi-particle self-consistent $GW$ (QS$GW$) methods, with and without vertex contributions (ladder diagrams) and examine the effects of spin-orbit coupling and the orbital composition of the bands.  The Bethe-Salpeter equation (BSE) method is used to study the optical response and the band origin of the low lying excitons is determined.
		The gradual transformation from the puckered $\alpha$-phase to the flat honeycomb structure is studied
		under biaxial strain and the evolution of the band structure and optical response is described in terms of band inversions of bands of different orbital character.
		
	\end{abstract}
	
	
	\maketitle
	
	\section{Introduction}
	Among two-dimensional monolayer materials, the elemental systems present a special interest. As we move from graphene \cite{Novoselov2004,castro2009} to silicene \cite{OUGHADDOU2015,KARA20121} and germanene \cite{Cahangirov2009} in group IV, the ideal flat honeycomb network gradually gives way to a buckled structure. In group-V elements—phosphorus, arsenic, antimony, and bismuth—new trends emerge.
	
	First, these elements carry one additional valence electron compared to group-IV. In a hypothetical flat honeycomb geometry, the Fermi level is therefore not pinned at a Dirac crossing but instead lies between a $p_z$-derived Dirac cone and another Dirac-like crossing dominated by the in-plane $p_x,p_y$ orbitals. Second, moving down the column increases relativistic effects: the $s$ and $p$ states separate more strongly, and spin–orbit coupling (SOC) becomes increasingly important. Third, unlike group-IV monolayers where buckling essentially always dominates, the group-V elements exhibit a competition between two structural motifs: the simple buckled ($\beta$) phase and the puckered ($\alpha$) phase, the latter being structurally analogous to black phosphorus \cite{Xia2019}.
	
	Previous studies have examined the evolution from the flat honeycomb structure to the buckled phase and reported interesting topological aspects, including band inversions and the movement and annihilation of Dirac crossings \cite{Radha2020,Radha20b,Radha20c,Radha21Topo}. In the present work, we focus on the puckered phase of arsenic monolayers with the corresponding crystal structure and Brillouin zone shown in Fig. \ref{fig:StrucView}. We present its electronic band structure, quasiparticle corrections, optical properties, and the evolution of the electronic structure during the structural transformation toward the flat honeycomb phase.
	
	\begin{figure}[htp]
		\centering
		
		\begin{minipage}[t]{0.48\linewidth}
			\centering
			\sidesubfloat[]{
				\includegraphics[width=\linewidth,trim=0cm 0cm 0cm 0cm]{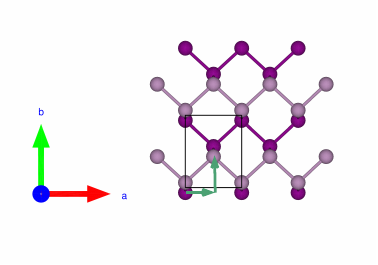}
			}\\[-4pt]
			\sidesubfloat[]{
				\includegraphics[width=\linewidth,trim=5.8cm 0cm 0cm 5cm]{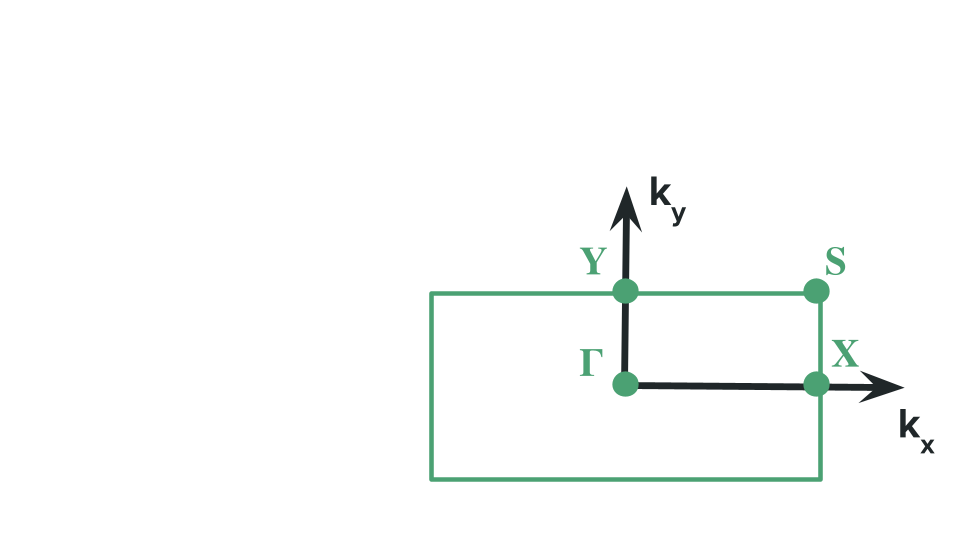}
			}
		\end{minipage}
		\hfill
		\begin{minipage}[t]{0.48\linewidth}
			\centering
			\sidesubfloat[]{
				\includegraphics[width=\linewidth,trim=0cm 0cm 0cm 0cm]{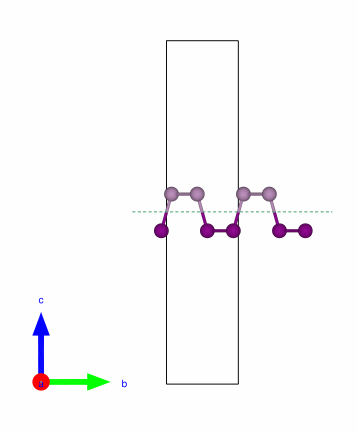}
			}
		\end{minipage}
		
		\caption{Crystal structure of $\alpha$-arsenene shown from the top view (a) and
			the side view (c), along with the corresponding reciprocal-space
			Brillouin zone and high-symmetry points (b). In panels (a) and (c), the
			green arrows and dashed line indicate a glide-mirror symmetry operation.
			The operation consists of a non-primitive translation followed by a
			mirror reflection, where the mirror plane passes halfway between the two
			sublattices, as illustrated by the dashed line in (c).
			The Cartesian axes are defined as $x \parallel a$, $y \parallel b$, and $z \parallel c$, where $y$ corresponds to the puckering direction and $x$ lies along the shorter in-plane lattice vector.
		}
		\label{fig:StrucView}
	\end{figure}
	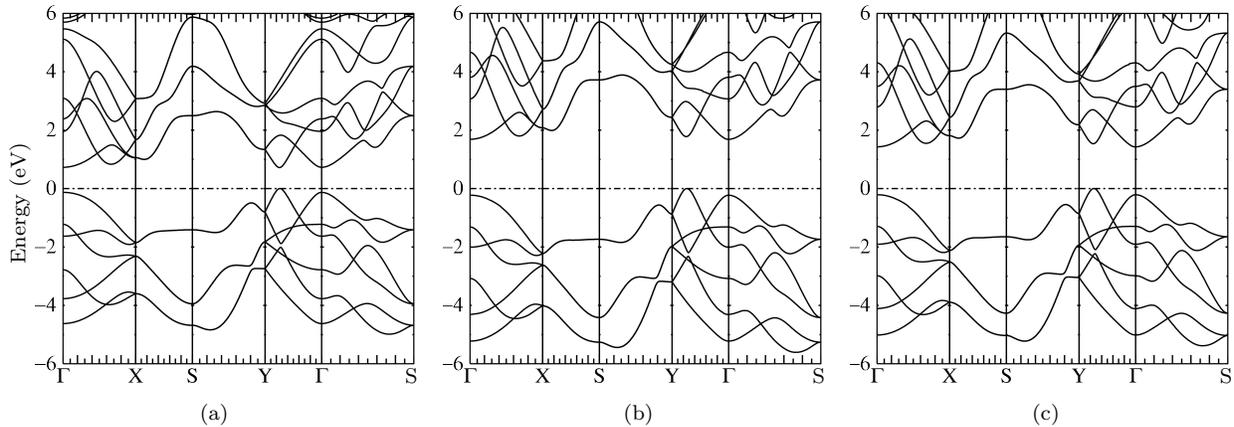
\begin{figure*}[t]
		\centering	
		\sidesubfloat[]{
			\begin{tikzpicture}
				\node[anchor=south west,inner sep=0] (img) at (0,0)
				{\includegraphics[width=0.29\linewidth,
					trim=4cm 4cm 2cm 8.5cm,clip]{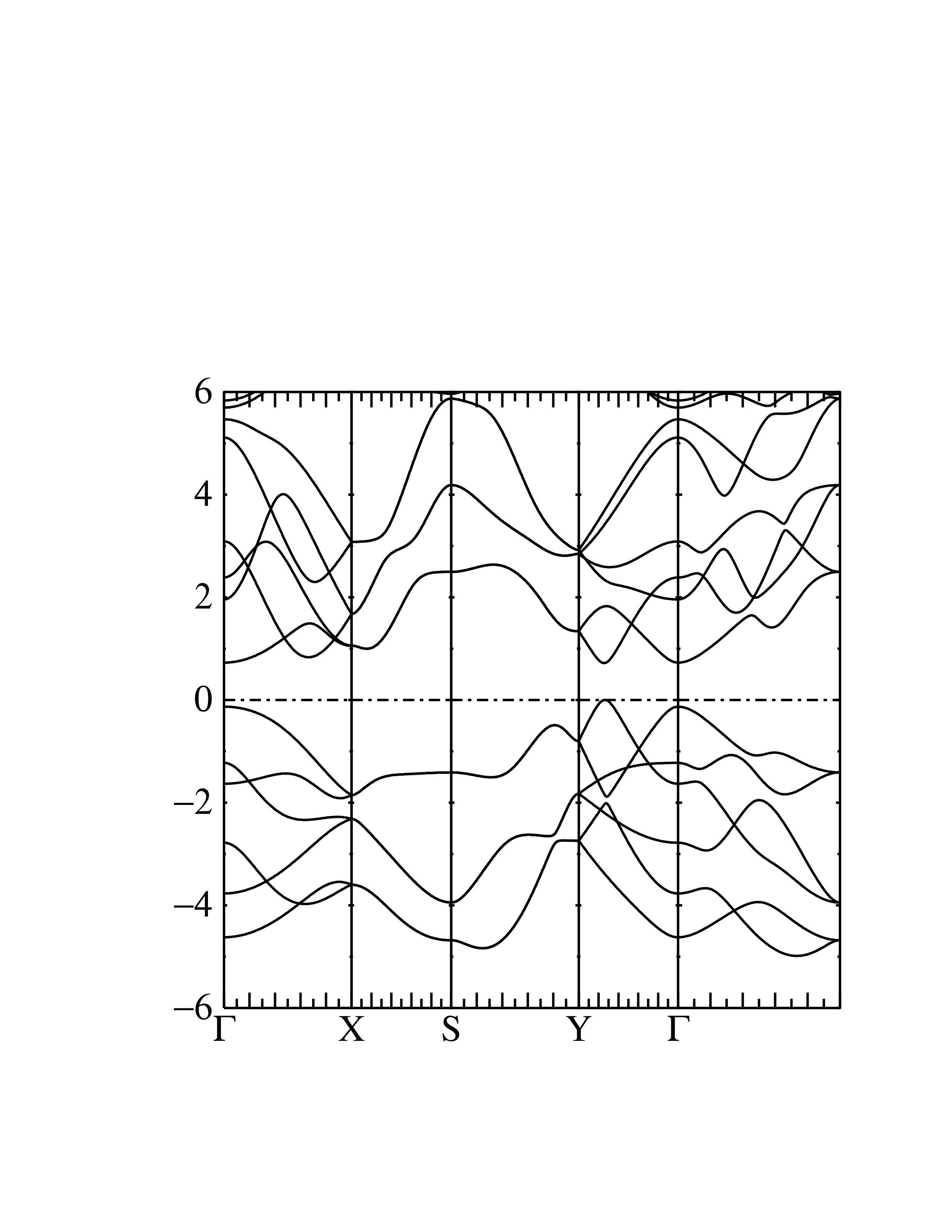}};
				
				\node[font=\footnotesize] at (5,0.2) {S};
				
				\node[rotate=90,font=\footnotesize] at (-0.2,2.5) {Energy (eV)};
			\end{tikzpicture}
		}%
		\sidesubfloat[]{
			\begin{tikzpicture}
				\node[anchor=south west,inner sep=0] (img) at (0,0)
				{\includegraphics[width=0.29\linewidth,
					trim=4cm 4cm 2cm 8.5cm,clip]{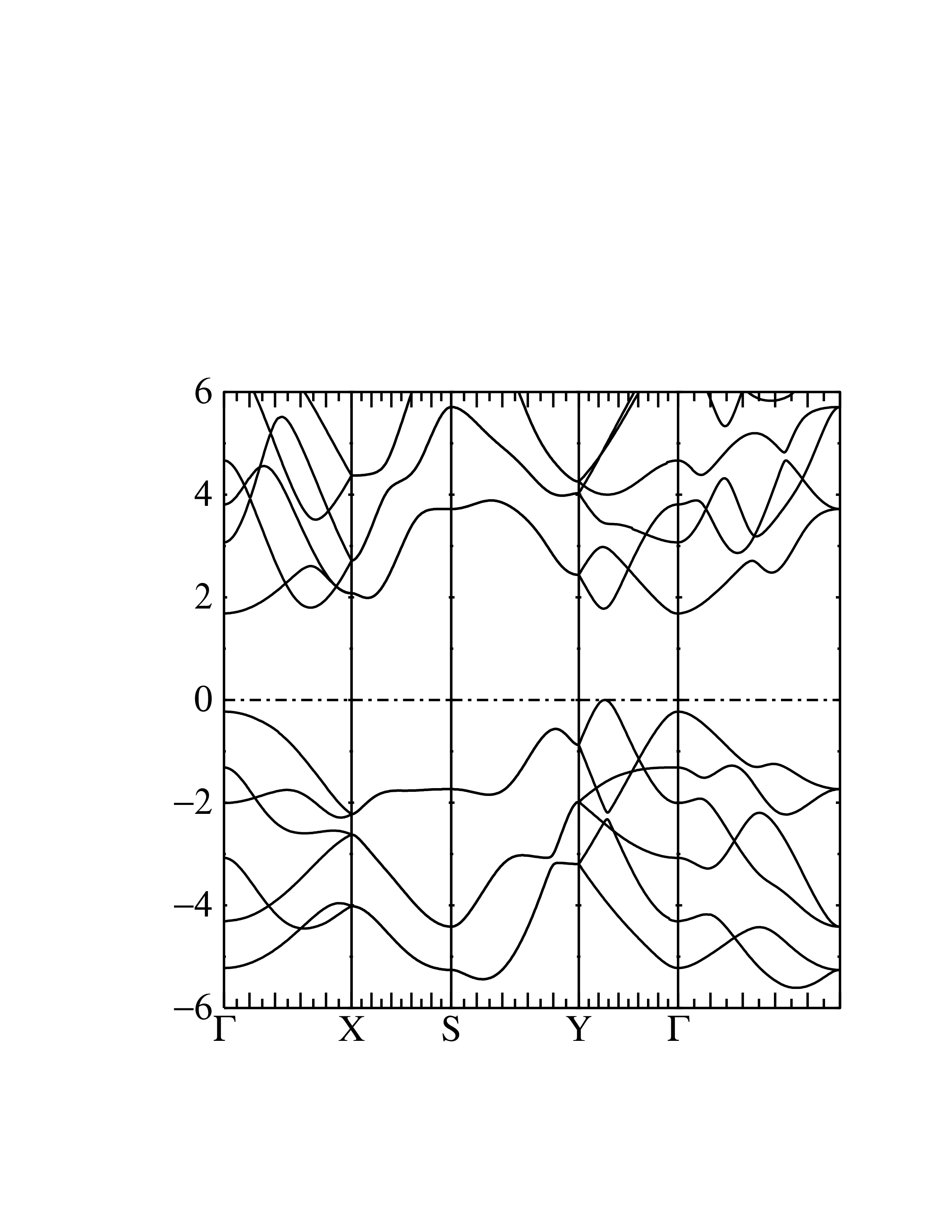}};
				
				\node[font=\footnotesize] at (5,0.2) {S};
			\end{tikzpicture}
		}%
		\sidesubfloat[]{
			\begin{tikzpicture}
				\node[anchor=south west,inner sep=0] (img) at (0,0)
				{\includegraphics[width=0.29\linewidth,
					trim=4cm 4cm 2cm 8.5cm,clip]{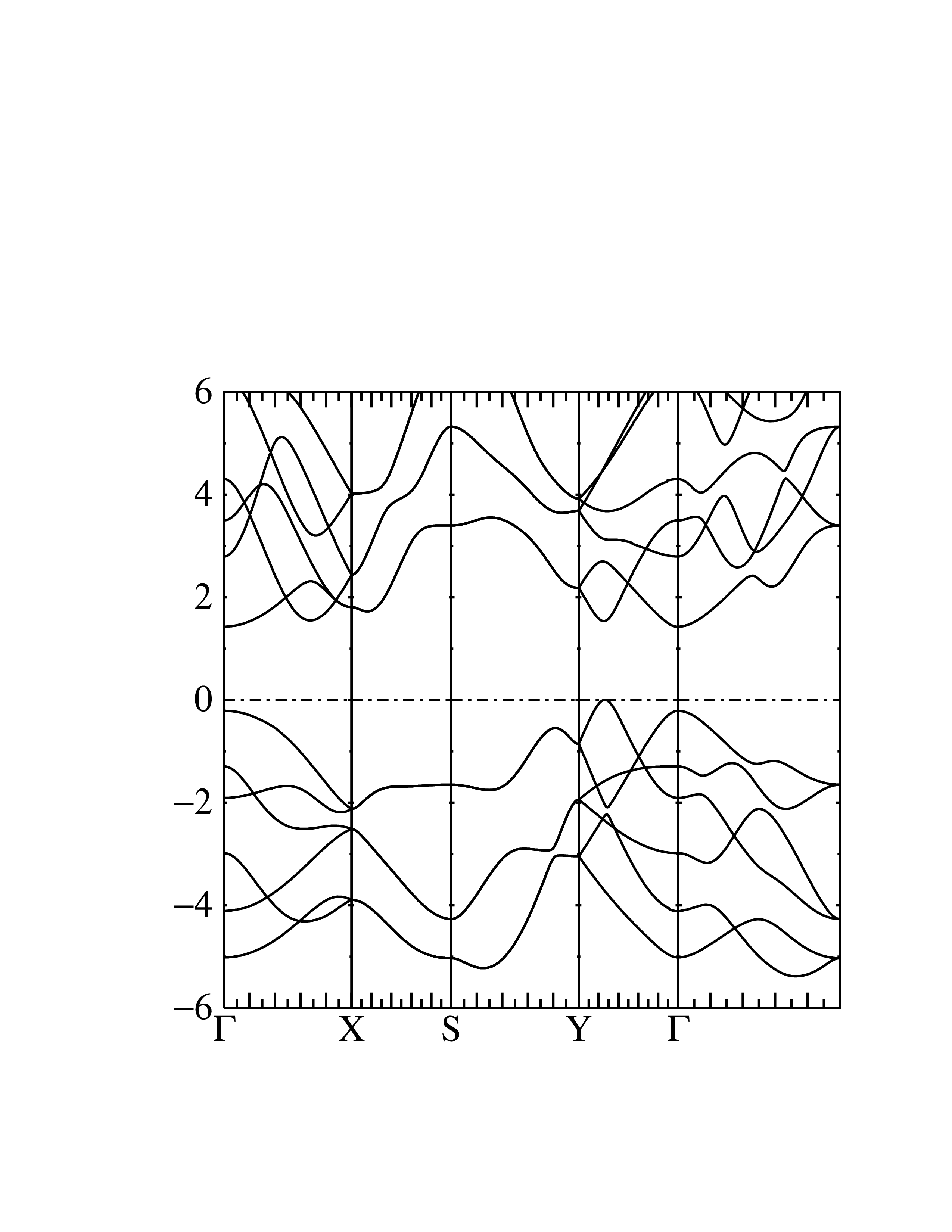}};
				
				\node[font=\footnotesize] at (5,0.2) {S};
			\end{tikzpicture}
		}	
		\caption{Band structure of the relaxed puckered phase calculated using (a) LDA, (b) QS$GW$, and (c) QS$G\hat{W}$. The $y$-axis shows energy (eV), and the $x$-axis follows the chosen high-symmetry path.}
		\label{fig:AsBands}
		
	\end{figure*}

	\section{Computational Methods}
	The band structure calculations in this work were done using the full-potential linearized muffin-tin orbital (FP-LMTO) method \cite{Methfessel2000,Kotani10} as implemented in the {\sc Questaal} suite of codes\cite{questaal}. This package implements both density functional theory (DFT) and many-body perturbation theory (MBPT) approaches.  In particular we use the quasiparticle self-consistent  (QS$GW$) version \cite{MvSQSGWprl,Kotani07} of Hedin's \cite{Hedin65,Hedin69} $GW$-method, where $G$ stands for the one-particle Green's function and $W$ for the screened Coulomb interaction, and  the Bethe-Salpeter Equation (BSE) method for calculating macroscopic dielectric function for optical properties \cite{1988StrinatiLaRivista,Hanke78,Onida02} as recently implemented with the LMTO and mixed-product-basis (MPB) set for two particle quantities by  Cunningham \etal \cite{Cunningham18,Cunningham23}.
	The latter is also used in an improved version of the QS$GW$ method with vertex corrections for the polarization propagator, representing electron-hole interactions via a ${\bf q}$-dependent BSE approach, which has been named QS$G\hat{W}$ or also QS$GW^{BSE}$. This approach avoids the systematic overestimate by about 20\%  of the self-energy $\Sigma$ and hence of the gaps in the original QS$GW$ approach which uses the Random Phase Approximation (RPA) for the polarization propagator, due to the corresponding underestimate of the screening dielectric constant.
	
	For the transition path between the puckered ($\alpha$) phase and the planar honeycomb structure, we employed {\sc Quantum Espresso} (QE). QE provides a straightforward and reliable approach for fully relaxing both internal atomic coordinates and lattice vectors by minimizing forces and the stress tensor. We used the projector augmented-wave (PAW) method \cite{PAW} with a $31\times21\times2$ $k$-mesh.
	For the QE relaxations, we used the following convergence parameters:
	\texttt{etot\_conv\_thr = 4.0d-5} for the total-energy threshold,
	\texttt{forc\_conv\_thr = 1.0d-4} for the force convergence,
	and \texttt{conv\_thr = 8.0d-10} for the electronic self-consistency.
	The relaxation was performed with the constraint
	\texttt{cell\_dofree = 'epitaxial\_ab'}, which allows only the in-plane
	lattice parameters to change while keeping the out-of-plane dimension fixed.
	For the plane-wave basis, we used a wavefunction cutoff of
	\texttt{ecutwfc = 70~Ry} and a charge-density cutoff of
	\texttt{ecutrho = 460~Ry}, consistent with the recommended values of the
	PAW pseudopotential.
	
	For the Questaal calculations, the basis consisted of a double set of smoothed Hankel envelope functions $spdf-spd$ with optimized decay lengths $\kappa$ and smoothing radii, with augmentation cutoff $l_\text{max}=4$. The $k$-mesh for QS$GW$ was $13\times10\times2$. The self-energy was represented with a cutoff of 2 Ry, above which it was replaced by an averaged diagonal value. The MPB used an angular-momentum cutoff $l=4$, GCUTX$=1.3~\text{Ry}^{1/2}$ for the interstitial product basis, and GCUTB$=1.6~\text{Ry}^{1/2}$ for the interstitial muffin-tin basis. To simulate the isolated monolayer limit, a vacuum region of approximately 23~\AA{} was employed. In the QS$G\hat{W}$ calculations, 6 valence and 6 conduction bands were included in the active space of the BSE Hamiltonian.
	
	\section{Results}
	
	\subsection{Quasiparticle Band Structures}
	
	Fig.~\ref{fig:AsBands} shows the band structure of the relaxed puckered phase obtained using LDA, QS$GW$, and QS$G\hat{W}$. For the DFT starting point, structural parameters were taken from the Materials Project \cite{MaterialsProject,MPas} and further relaxed using QE. The in-plane lattice constants are $a = 3.69$~\AA{} and $b = 4.72$~\AA{}, and a vacuum spacing of 22.5~\AA{} was used along the out-of-plane direction to avoid interactions between periodic images.
	In LDA, the gap is direct at an intermediate point between $\Gamma$ and Y. After including quasiparticle corrections, both QS$GW$ and QS$G\hat{W}$ predict an indirect gap, with the VBM located between $\Gamma$ and Y and the CBM at $\Gamma$. The lowest conduction band also displays shallow local minima along $\Gamma$–X and $\Gamma$–Y, lying close in energy to the CBM. These features are expected to be sensitive to structural parameters and may become relevant when tracing the evolution toward the planar phase.
	
\begin{figure}[htp]
	\centering
	\begin{tikzpicture}
		\node[anchor=south west,inner sep=0] (img) at (0,0)
		{\includegraphics[width=.8\linewidth]{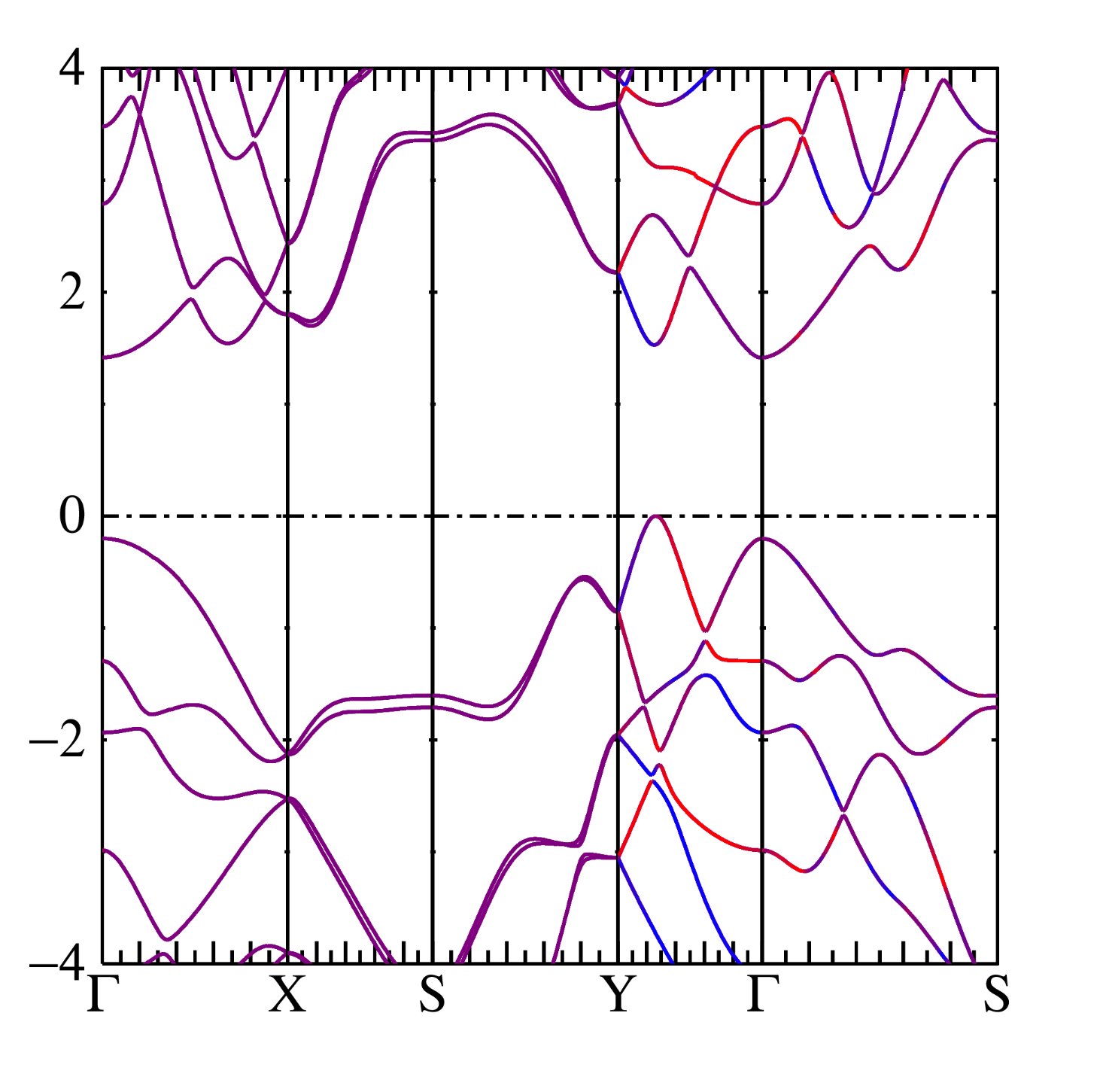}};
		\node[rotate=90,font=\small] at (-0.05,3.5) {Energy (eV)};
	\end{tikzpicture}
	\caption{Band structure of $\alpha$-arsenene including SOC. Spin--orbit coupling
		mixes the spin character of the bands; blue and red colors indicate
		spin-up and spin-down contributions, respectively, with intermediate
		colors corresponding to mixed character.}
	\label{fig:bandsoc}	
\end{figure}

	Along the Brillouin-zone edge (X–S–Y), the bands are doubly degenerate in the calculations without spin-orbit coupling (SOC). This degeneracy arises from the non-symmorphic nature of the space group.
	The band structure with SOC for the final QS$G\hat{W}$ case is shown in Fig. \ref{fig:bandsoc}. After including SOC, these degeneracies are lifted everywhere except exactly at X and Y, leaving behind symmetry-protected two-dimensional Dirac points. These SOC-induced splittings differ from the Dirac physics in graphene, since the states here are explicitly spin–orbit coupled.
	SOC is added \textit{a posteriori} to the QS$GW$ or QS$G\hat{W}$ Hamiltonian; the self-energy itself is computed without a fully four-component relativistic treatment.
	
\begin{figure}[htp]
		\centering
		\begin{tikzpicture}
			\node[anchor=south west,inner sep=0] (img1) at (0,0)
			{(a)\includegraphics[width=.8\linewidth,
				trim=4cm 4cm 2cm 8.5cm,clip]{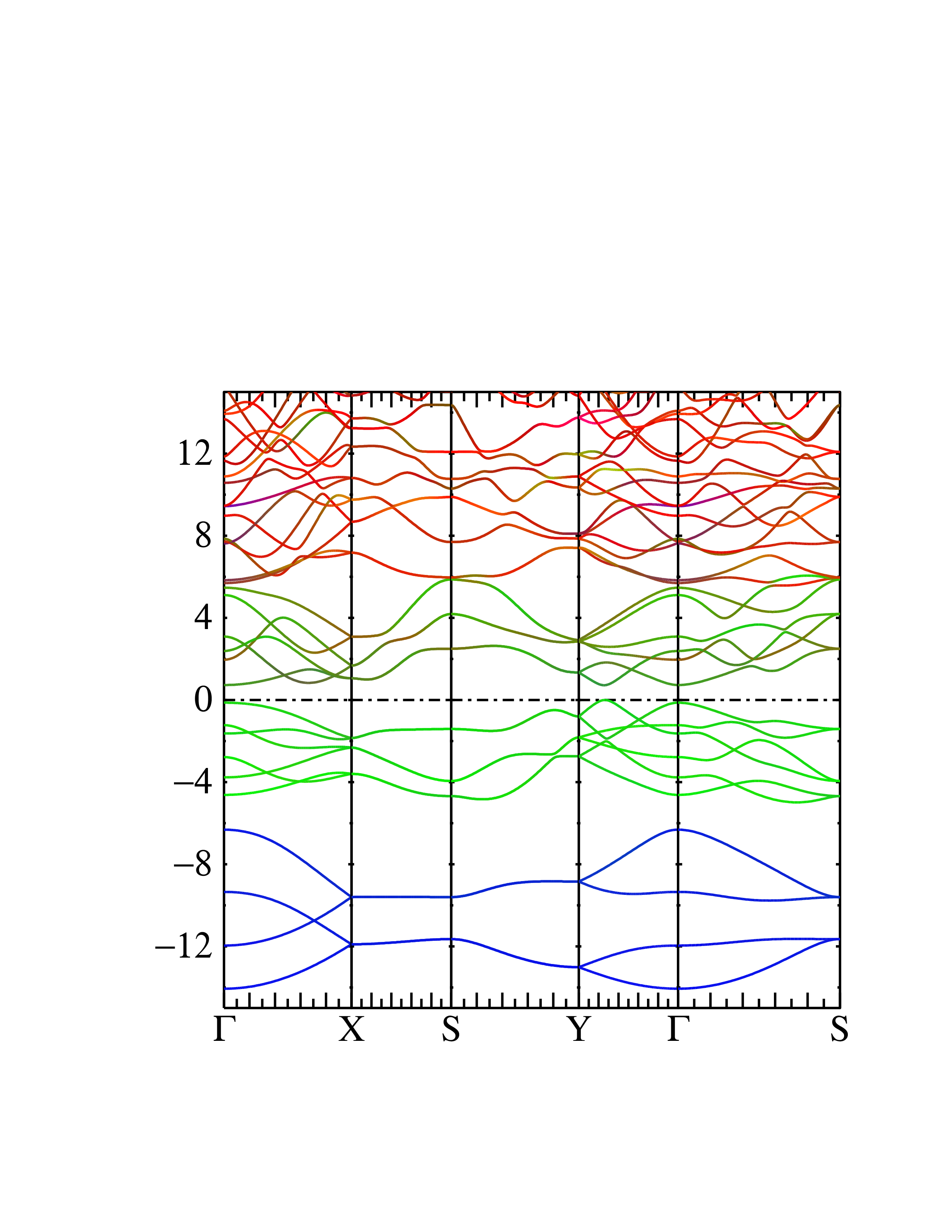}};
			\node[rotate=90,font=\small] at (-0.3,3.5) {Energy (eV)};
			\node[anchor=north west,inner sep=0] (img2) at (img1.south west)
			{(b)\includegraphics[width=.8\linewidth,
				trim=4cm 4cm 2cm 8.5cm,clip]{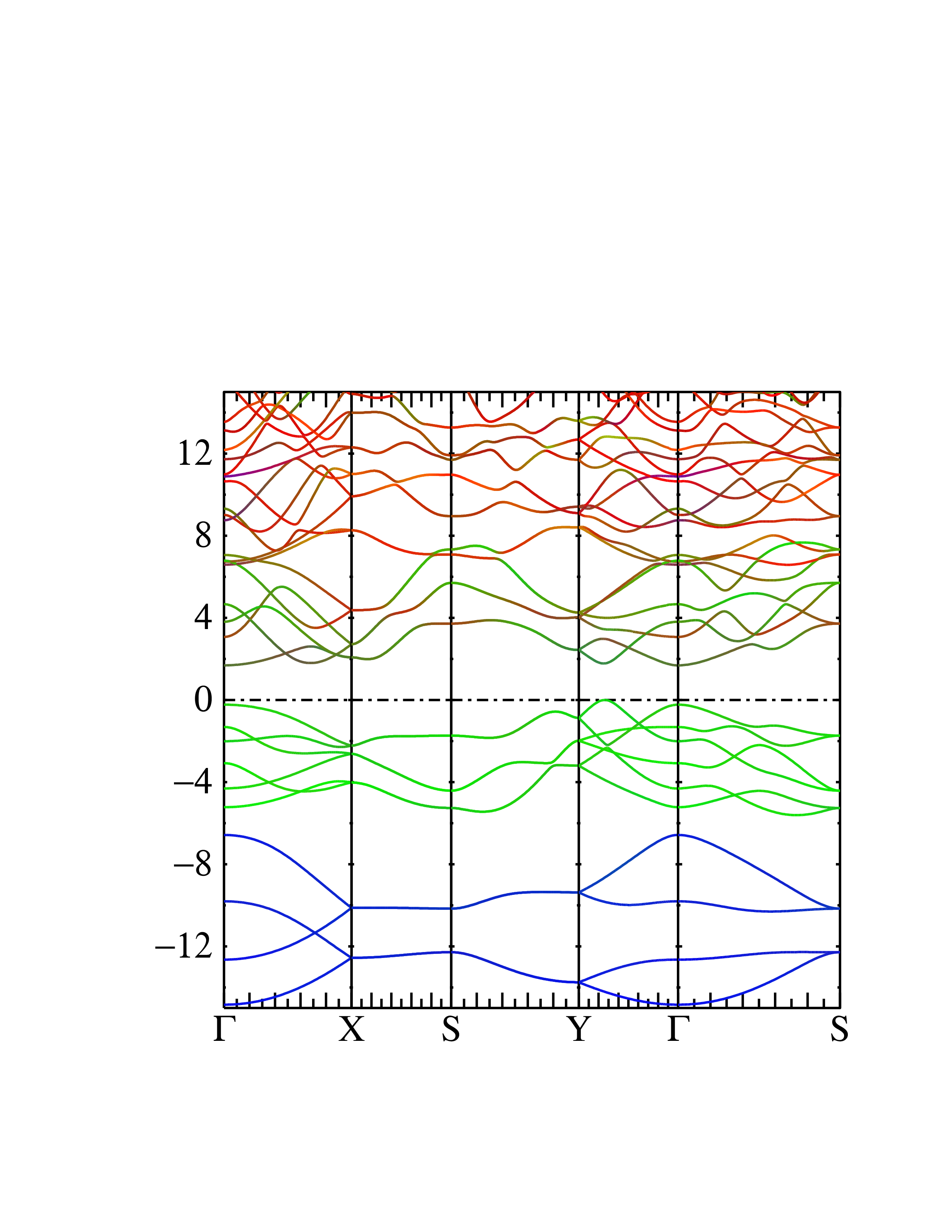}};
			\node[rotate=90,font=\small] at (-0.3,-3.5) {Energy (eV)};
		\end{tikzpicture}
		\caption{Orbital-resolved band structures from (a) LDA and (b) QS$GW$ calculations. Colors denote $s$ (blue), $p$ (green), and $d$ (red) character.}
		\label{fig:AsBandsOrbitals}
	\end{figure}

	To analyze the band dispersion in more detail, Fig.~\ref{fig:AsBandsOrbitals} shows the $s$, $p$, and $d$ orbital contributions, shown as blue, green and red respectively.
	The lowest four bands correspond to As-$s$ states, well separated from the rest. The six valence bands nearest the Fermi level arise from bonding combinations of the twelve As-$p$ orbitals in the cell, while the corresponding antibonding combinations form the lowest six conduction bands. These conduction states already show a notable hybridization with the As-$d$ orbitals, which dominate above approximately 7 eV. The degree of $p$–$d$ hybridization is strongly $\mathbf{k}$-dependent and particularly enhanced near the S point. QS$GW$ amplifies this effect compared to LDA and swaps the $p$ and $d$ character of the bands at certain points in the k-space. Speficially, the lowest conduction band at $S$ has a strong $d$ character but becomes more $p$-like as we move away topard $X$ or $Y$. The second conduction band has the opposite behavior, it is $p$-like at $S$ and becomes $d$-like at some point between $S$ and $X$ (or $Y$), indicating a band-character inversion near this point. Closer to $X$, the $d$-character switches to the third conduction band. 
	
	This orbital analysis will be important in identifying possible band inversions as the structure is gradually transformed toward the planar honeycomb geometry, which will be examined in the next section.
	
	
	\subsection{Optical dielectric function and excitons}
	
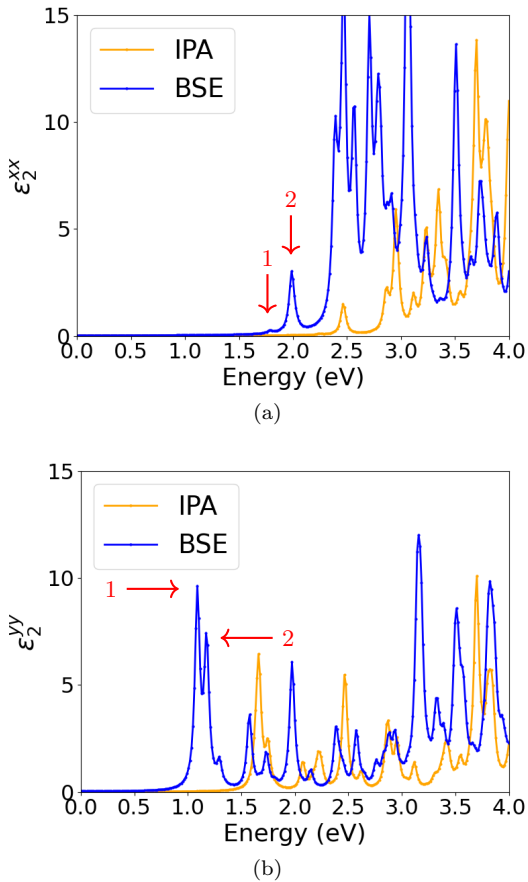
\begin{figure}[htp]
		\centering
		
		\sidesubfloat[]{
			\begin{tikzpicture}
				\node[anchor=south west,inner sep=0] (img) at (0,0)
				{\includegraphics[width=.8\linewidth,
					trim=.3cm .3cm .3cm .1cm,clip]{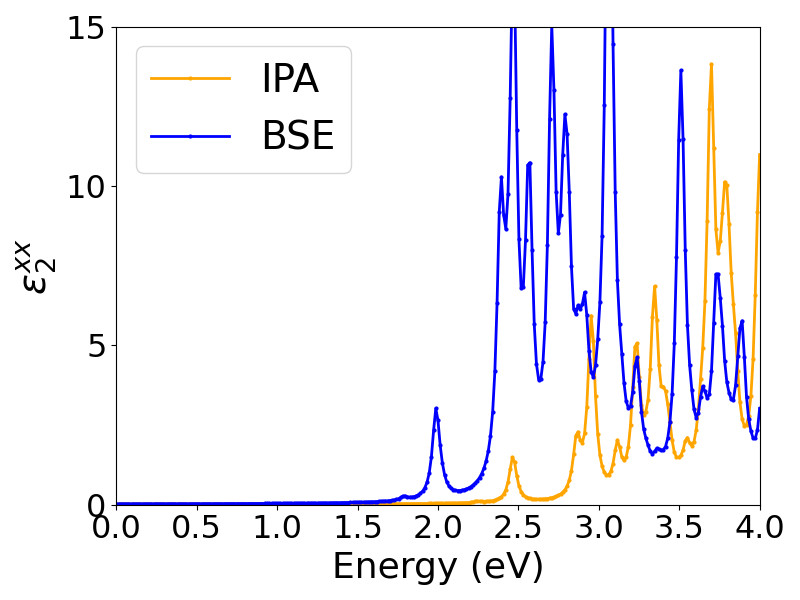}};
				
				\begin{scope}[x={(img.south east)},y={(img.north west)}]
					\draw[->, thick, red] (0.5,0.3) -- (0.5,0.20);
					\node at (0.5,0.34) {\textcolor{red}{1}};
					
					\draw[->, thick, red] (0.545,0.45) -- (0.545,0.35);
					\node at (0.545,0.49) {\textcolor{red}{2}};
				\end{scope}
			\end{tikzpicture}
		}\\
		
		\sidesubfloat[]{
			\begin{tikzpicture}
				\node[anchor=south west,inner sep=0] (img) at (0,0)
				{\includegraphics[width=.8\linewidth,
					trim=.3cm .3cm .3cm .1cm,clip]{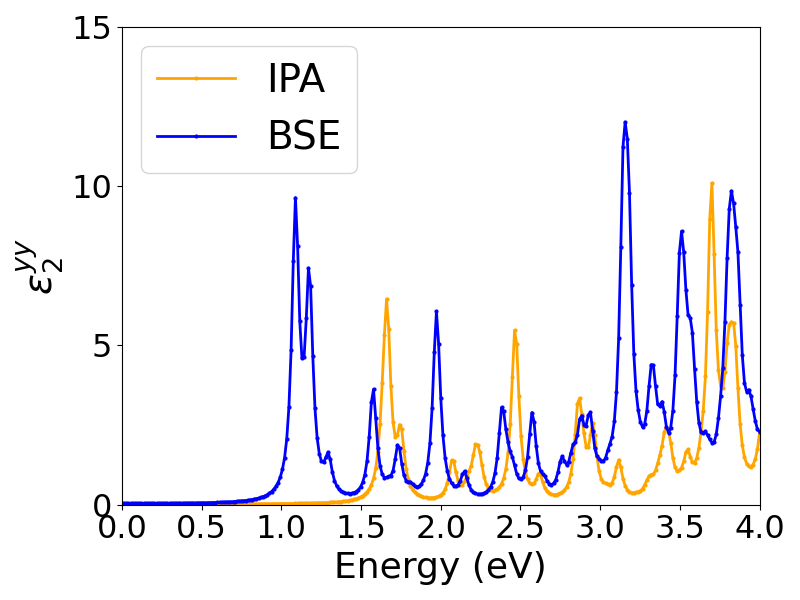}};
				
				\begin{scope}[x={(img.south east)},y={(img.north west)}]
					\draw[->, thick, red] (0.23,0.66) -- (0.33,0.66);
					\node at (0.2,0.66) {\textcolor{red}{1}};
					
					\draw[->, thick, red] (0.51,0.535) -- (0.41,0.535);
					\node at (0.54,0.535) {\textcolor{red}{2}};
				\end{scope}
			\end{tikzpicture}
		}
		
		\caption{Imaginary part of the macroscopic dielectric function for
			(a) $x$-polarized and (b) $y$-polarized light. Orange:
			independent-particle approximation (IPA). Blue: Bethe--Salpeter
			equation (BSE). The first and second bright excitons are labeled as 1 and 2, respectively.}
		\label{fig:eps2}
	\end{figure}

\begin{figure}[htp]
		\centering
		\sidesubfloat[]{
			\includegraphics[width=.8\linewidth,
			trim=0cm 0cm 0cm 0cm,clip]{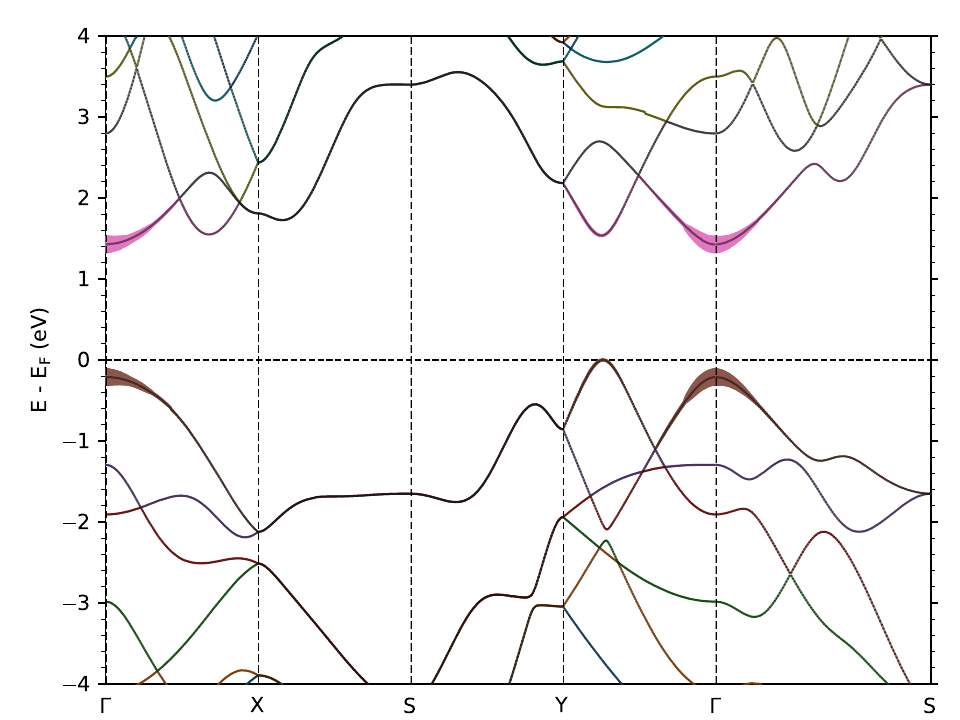}
		}\\
		\sidesubfloat[]{
			\includegraphics[width=.8\linewidth,
			trim=0cm 0cm 0cm 0cm,clip]{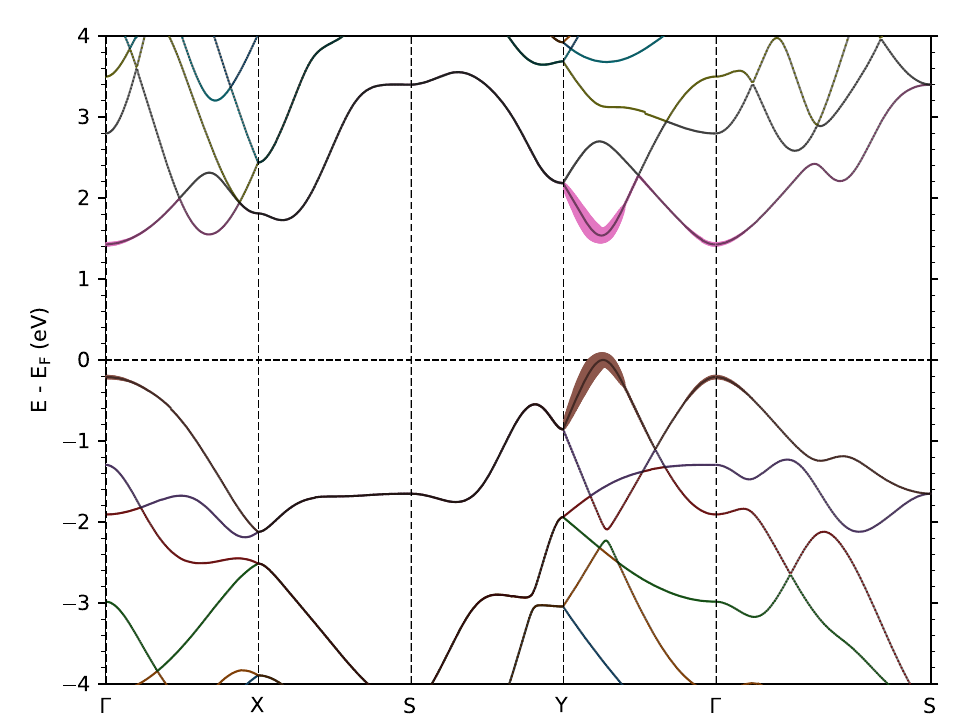}
		}
		\caption{$\mathbf{k}$-space distributions of the first two bright
			excitons with $y$ polarization. (a) First exciton, at 1.089~eV, mainly near
			$\Gamma$. (b) Second exciton, at 1.175~eV, concentrated along the $\Gamma$–Y
			path.}
		\label{fig:kdisty}
	\end{figure}

\begin{figure}[htp]
		\centering
		\sidesubfloat[]{
			\includegraphics[width=.8\linewidth,
			trim=0cm 0cm 0cm 0cm,clip]{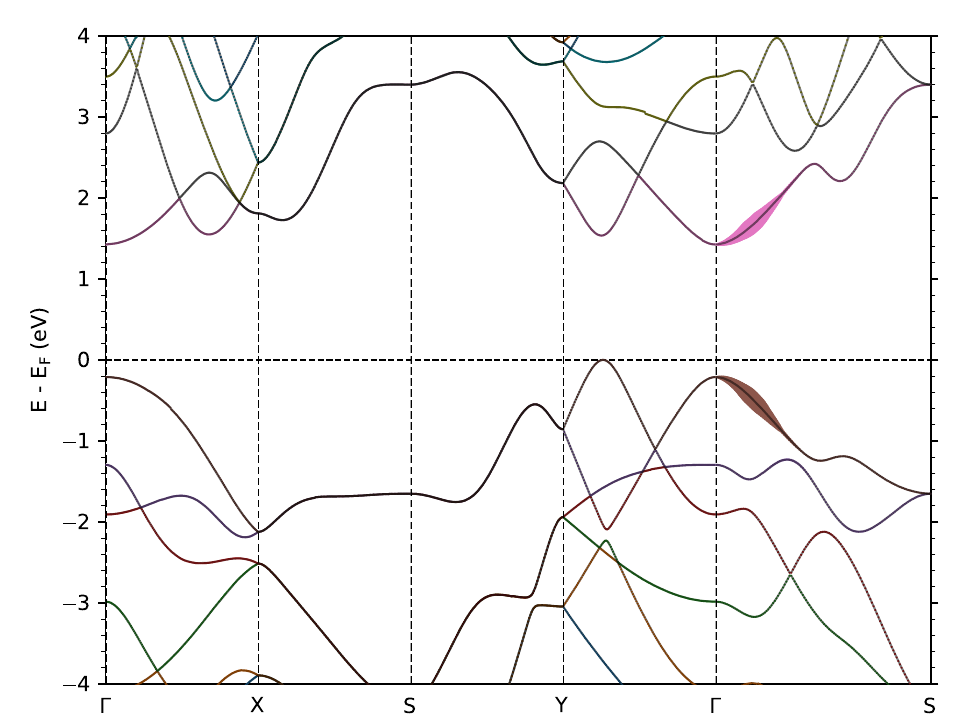}
		}\\
		\sidesubfloat[]{
			\includegraphics[width=.8\linewidth,
			trim=0cm 0cm 0cm 0cm,clip]{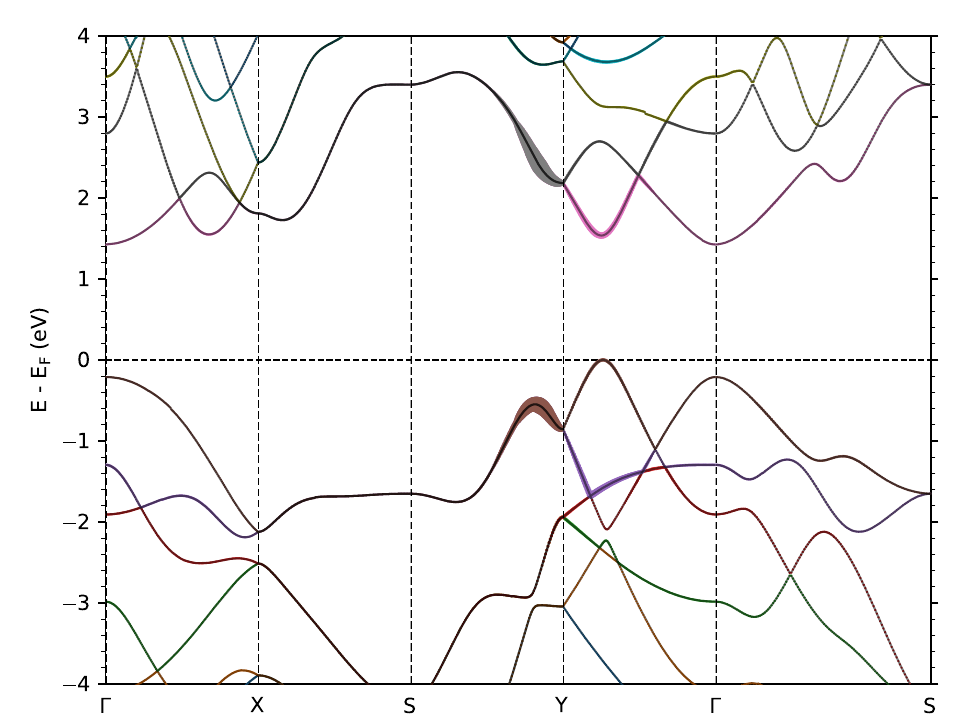}
			\label{fig:kdistxb}
		}
		\caption{$\mathbf{k}$-space distributions of the first two bright
			excitons for $x$ polarization. (a) Lowest bright exciton at 1.787~eV. (b) Second
			bright exciton, at 1.988~eV, plotted with weights scaled by 100 for visibility.}
		\label{fig:kdistx}
	\end{figure}

	Fig.~\ref{fig:eps2} shows the imaginary part of the macroscopic
	dielectric function for light polarized along the two in-plane
	directions. The independent-particle approximation (IPA) spectrum is
	compared with the corresponding BSE result,
	where electron--hole interactions and local-field effects are explicitly
	included. As expected for the puckered structure, the optical response
	is strongly anisotropic: the $x$ and $y$ polarizations show different
	onset energies and different distributions of oscillator strength.
	In both polarizations the BSE spectrum is red-shifted relative to the
	IPA result, reflecting the binding of electron--hole pairs. The first bright exciton appears near
	$1.8$~eV for $x$ polarization and around $1.1$~eV for $y$ polarization.
	The oscillator strength is noticeably weaker along $x$.
	To visualize the nature of the lowest excitons, we show in
	Figs.~\ref{fig:kdisty} and \ref{fig:kdistx} the $\mathbf{k}$-space
	distribution of the first two bright BSE states for $y$- and
	$x$-polarized light, respectively. In the present BSE implementation,
	only vertical (same-$\mathbf{k}$) transitions between valence and
	conduction bands are included.
	
	For $y$ polarization (Fig.~\ref{fig:kdisty}), both excitons originate
	predominantly from transitions between the topmost valence band and the
	lowest conduction band. The lowest-energy exciton is concentrated around
	the $\Gamma$ point, where a small direct gap exists even though the
	global quasiparticle gap is indirect. The second bright exciton is
	localized along the $\Gamma$–Y direction. Its weight is asymmetric,
	with a stronger contribution closer to the Y point, consistent with the
	local direct gap along this path.
	For $x$ polarization (Fig.~\ref{fig:kdistx}), the lowest bright exciton
	also originates from transitions between the highest valence and lowest
	conduction bands, with a $\mathbf{k}$-space distribution along $\Gamma$--S direction. This reduced strength is
	consistent with the BSE spectrum, where the first peak in the $x$
	direction is small. The second bright exciton has its weight distributed
	over several bands and $\mathbf{k}$ points; along the high-symmetry
	paths, it is most prominent along the S--Y direction. For visibility, the
	weights plotted in Fig.~\ref{fig:kdistxb} are scaled by a factor of 100.
	Finally, the contributions of these two excitons along the S--$\Gamma$
	and S--Y lines show a termination at the $\Gamma$ and Y points,
	respectively.
	In addition to the bright excitons discussed above, the BSE eigenvalue
	spectrum also contains optically inactive (dark) states. These dark
	excitons correspond to eigenvectors with vanishing oscillator strength for the chosen polarization, which depends on a combination of optical matrix elements and exciton envelope functions. For relaxed puckered structure the first exciton is bright, however as we apply the strain, the first exciton becomes dark. More detail on these changes will be presented  in the next section. 
	
	\subsection{Structural Relaxation}
	To investigate the changes in band structure from the puckered to the flat phase, we first need to find a realistic way to convert the structure under strain. 
	Structural relaxations were performed under both uniaxial and biaxial
	in-plane strain using the QE package. For small uniaxial strains,
	our results are consistent with the findings of Kamal and Ezawa
	\cite{Kamal2015}. Extending the range to larger strains, however does not lead to the flat hexagonal structure. Only
	biaxial strain is capable of fully flattening the layer into a
	honeycomb-like configuration. Under uniaxial strain, applied along
	either the $x$ or $y$ direction, the material behaves normally: we do
	not observe auxetic response, meaning that stretching the structure
	along one direction causes contraction along the perpendicular direction
	(i.e. a positive in-plane Poisson ratio).  However, examining the flat hexagonal structure in the corresponding doubled supercell of the puckered structure, 
	we see that we need to have a tensile strain in both in-plane directions, which can never be achieved with uniaxial strain 
	 Therefore we now focus on biaxial strain. 
	
	For the biaxial case, we interpolated the in-plane lattice constants
	$a$ and $b$ between their values in the relaxed puckered phase and those
	of the flat configuration. To maintain a consistent cell shape
	throughout this interpolation, the flat phase was represented using a
	doubled orthorhombic unit cell rather than its primitive hexagonal
	cell. The lattice constants evolve from $a = 3.69$~\AA~to
	$4.37$~\AA~and from $b = 4.72$~\AA~to $7.56$~\AA~, sampled at 33 intermediate points.
	
\begin{figure}[htp]
		\centering
		\sidesubfloat[]{
			\includegraphics[width=.45\linewidth]{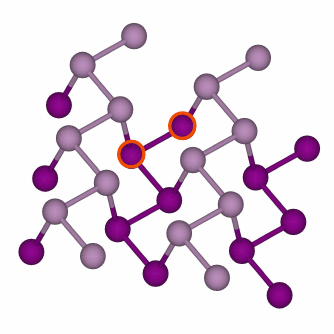}
		}%
		\sidesubfloat[]{
			\includegraphics[width=.45\linewidth]{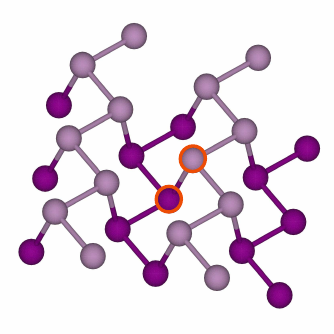}
		}\\
		\sidesubfloat[]{
			\includegraphics[width=.45\linewidth]{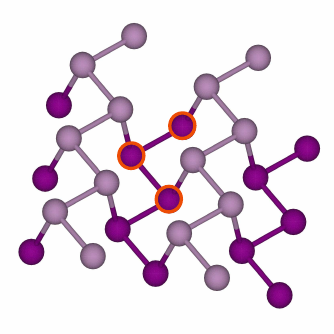}
		}%
		\sidesubfloat[]{
			\includegraphics[width=.45\linewidth]{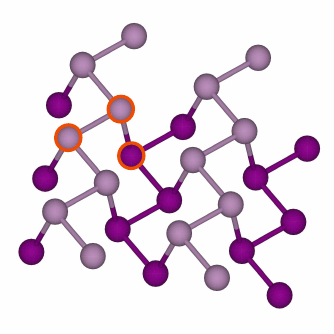}
		}
		\caption{Guide to the structural parameters that vary under strain, as shown in
			Fig.~\ref{fig:relax}. Orange circles indicate (a) the in-plane bond
			length $d_1$, (b) the out-of-plane bond length $d_2$, (c) the in-plane
			bond angle $\theta_1$, and (d) the out-of-plane bond angle $\theta_2$.
			Upon going from the relaxed puckered structure to the flat limit,
			$d_1$ changes from 2.509 to 2.521~\AA{}, $d_2$ from 2.490 to 2.523~\AA{},
			$\theta_1$ from $94.783^\circ$ to $120.057^\circ$, and $\theta_2$ from
			$100.367^\circ$ to $119.971^\circ$. The two shades of magenta distinguish the two sublayers separated along the out-of-plane direction.
		}
		\label{fig:StrucGuide}
	\end{figure}

\begin{figure*}[t]
		\centering
		\sidesubfloat[]{
			\includegraphics[width=0.27\linewidth]{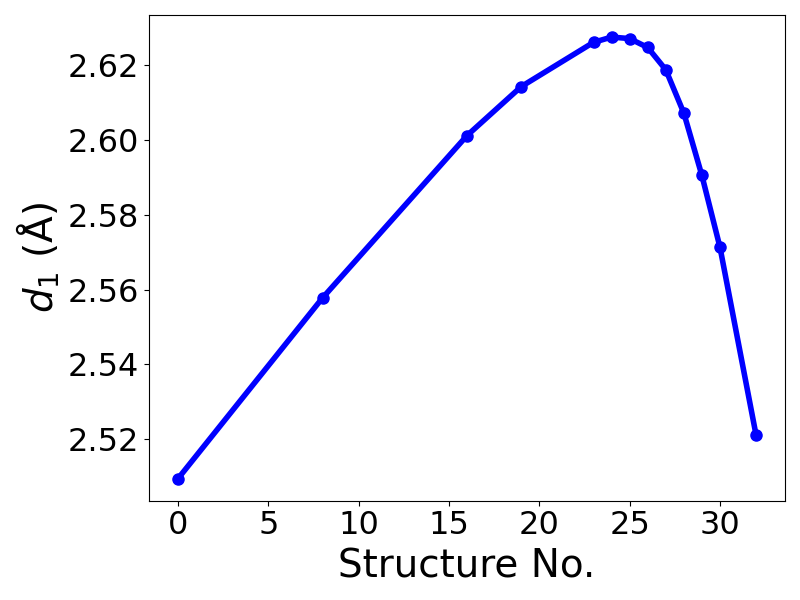}
		}
		\sidesubfloat[]{
			\includegraphics[width=0.27\linewidth]{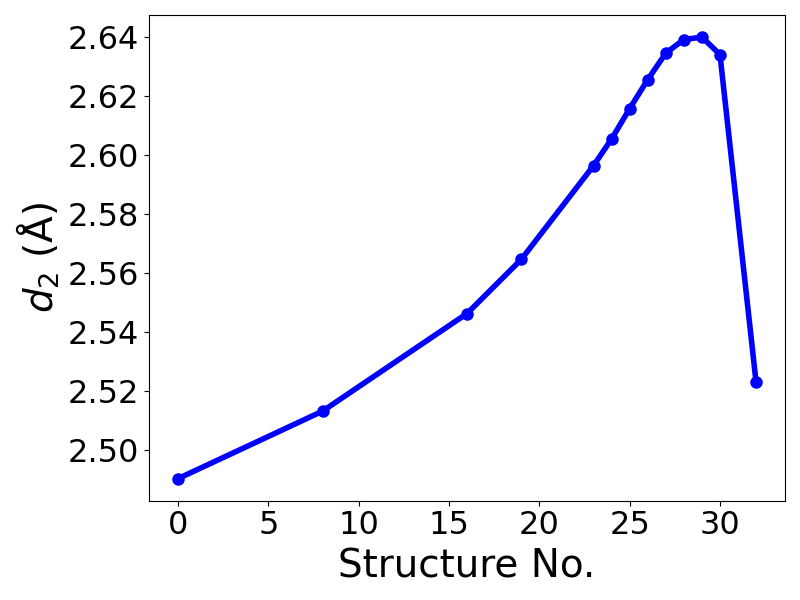}
		}
		\sidesubfloat[]{
			\includegraphics[width=0.27\linewidth]{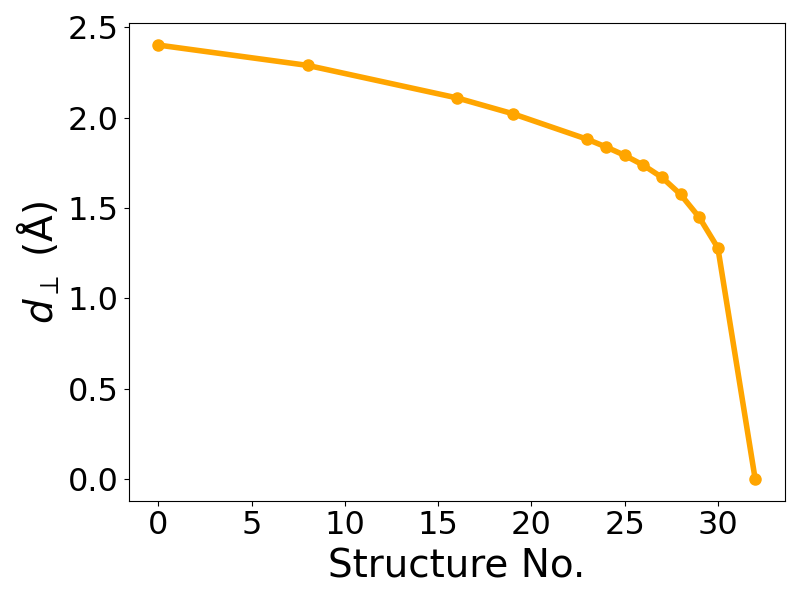}
		}\\
		\sidesubfloat[]{
			\includegraphics[width=0.27\linewidth]{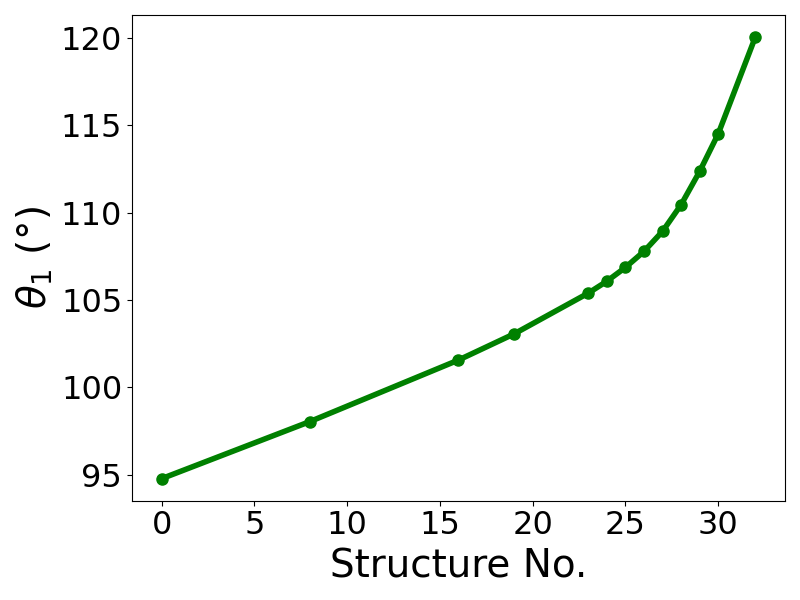}
		}
		\sidesubfloat[]{
			\includegraphics[width=0.27\linewidth]{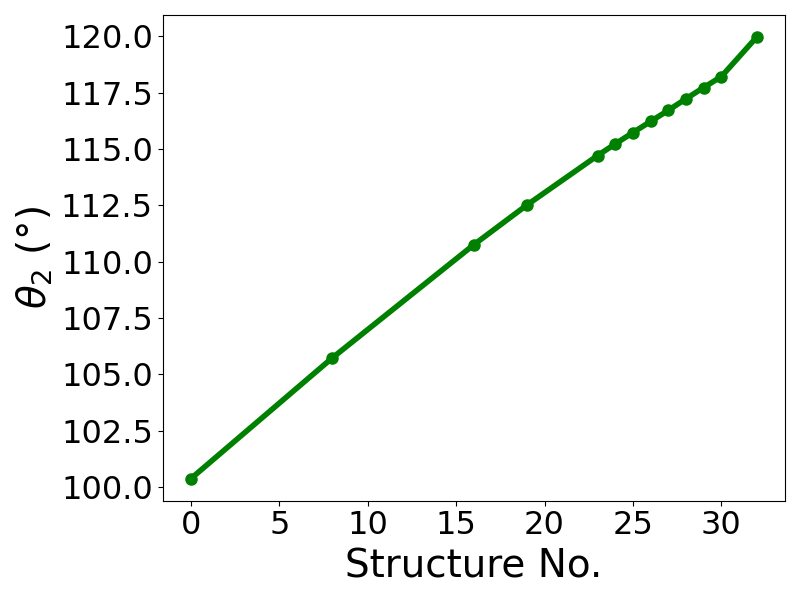}
		}
		\caption{Structural relaxation results under biaxial strain,
			showing (a) $d_1$, (b) $d_2$, (c) $d_\perp$, (d) $\theta_1$, and
			(e) $\theta_2$ for structures 0--32. Here, $d_1$ and $d_2$ denote
			in-plane and out-of-plane bond lengths, respectively, while
			$d_\perp$ is the vertical separation between the two sublattices.
			Under tensile strain, $d_1$ and $d_2$ increase non-linearly.
			Toward the end of the strain range they rapidly converge, and
			$d_\perp$ collapses to zero, signaling the transition to a flat
			geometry. The bond angles $\theta_1$ and $\theta_2$ increase
			steadily and approach $120^\circ$, consistent with the formation of
			a planar honeycomb structure.}
		\label{fig:relax}
	\end{figure*}

	In Fig~\ref{fig:StrucGuide}, we define the key structural parameters in the puckered structure which evolve as the biaxial strain flattens out the structure. Fig.~\ref{fig:relax}~summarizes the evolution of the structural
	parameters along the biaxial strain path. Up to approximately step~25,
	the parameters change smoothly and monotonically, though not linearly. The guide to these parameters in separately shown in Fig. \ref{fig:StrucGuide}.
	\definecolor{lightgray}{gray}{0.9}
	\begin{table*}[t]
		\centering
		\renewcommand{\arraystretch}{0.8}
		\caption{Applied biaxial strain values for each structure relative to the unstrained cell.}
		\label{tab:biaxial-strain}
		\begin{tabular}{c c c c c c c}
			\toprule
			\textbf{Structure No.} & \textbf{$a$ (\AA)} & \textbf{$b$ (\AA)} &
			\textbf{$\varepsilon_{xx}$ (\%)} & \textbf{$\varepsilon_{yy}$ (\%)} &
			\textbf{Avg. biaxial (\%)} & \textbf{Areal strain (\%)} \\
			\midrule
			
			\rowcolor{lightgray} 0  & 3.69 & 4.72 & 0.00 & 0.00 & 0.00 & 0.00 \\
			1  & 3.71 & 4.81 & 0.57 & 1.88 & 1.23 & 2.46 \\
			2  & 3.74 & 4.90 & 1.14 & 3.76 & 2.45 & 4.95 \\
			3  & 3.76 & 4.99 & 1.71 & 5.62 & 3.66 & 7.49 \\
			4  & 3.79 & 5.08 & 2.28 & 7.38 & 4.83 & 10.08 \\
			5  & 3.81 & 5.17 & 2.85 & 9.44 & 6.14 & 12.75 \\
			6  & 3.84 & 5.25 & 3.42 & 11.25 & 7.34 & 15.37 \\
			7  & 3.86 & 5.34 & 3.99 & 13.09 & 8.54 & 18.09 \\
			\rowcolor{lightgray} 8  & 3.89 & 5.43 & 4.56 & 14.99 & 9.78 & 20.89 \\
			9 & 3.91 & 5.52 & 5.13 & 16.93 & 11.03 & 23.71 \\
			10 & 3.94 & 5.61 & 5.70 & 18.86 & 12.28 & 26.63 \\
			11 & 3.96 & 5.70 & 6.27 & 20.79 & 13.53 & 29.60 \\
			12 & 3.99 & 5.79 & 6.84 & 22.66 & 14.75 & 32.58 \\
			13 & 4.01 & 5.88 & 7.41 & 24.61 & 16.01 & 35.66 \\
			14 & 4.04 & 5.97 & 7.98 & 26.54 & 17.26 & 38.75 \\
			15 & 4.06 & 6.05 & 8.55 & 28.36 & 18.46 & 41.76 \\
			\rowcolor{lightgray} 16 & 4.09 & 6.14 & 9.12 & 30.24 & 19.68 & 44.89 \\
			17 & 4.11 & 6.23 & 9.69 & 32.06 & 20.88 & 47.98 \\
			18 & 4.14 & 6.32 & 10.26 & 33.87 & 22.06 & 51.08 \\
			19 & 4.16 & 6.41 & 10.83 & 35.83 & 23.33 & 54.33 \\
			20 & 4.19 & 6.50 & 11.40 & 37.65 & 24.52 & 57.52 \\
			21 & 4.21 & 6.59 & 11.97 & 39.49 & 25.73 & 60.73 \\
			22 & 4.24 & 6.68 & 12.54 & 41.32 & 26.93 & 63.96 \\
			\rowcolor{lightgray} 23 & 4.26 & 6.77 & 13.11 & 43.16 & 28.13 & 67.19 \\
			24 & 4.29 & 6.85 & 13.68 & 44.98 & 29.33 & 70.42 \\
			\rowcolor{lightgray} 25 & 4.31 & 6.94 & 14.25 & 46.82 & 30.54 & 73.69 \\
			26 & 4.33 & 7.03 & 14.82 & 48.65 & 31.74 & 76.96 \\
			\rowcolor{lightgray} 27 & 4.36 & 7.12 & 15.39 & 50.48 & 32.94 & 80.24 \\
			\rowcolor{lightgray} 28 & 4.38 & 7.21 & 15.96 & 52.32 & 34.14 & 83.53 \\
			\rowcolor{lightgray} 29 & 4.41 & 7.30 & 16.53 & 54.15 & 35.34 & 86.82 \\
			\rowcolor{lightgray} 30 & 4.33 & 7.39 & 17.11 & 56.45 & 36.78 & 83.21 \\
			31 & 4.35 & 7.48 & 17.68 & 58.33 & 38.00 & 86.32 \\
			\rowcolor{lightgray} 32 & 4.37 & 7.56 & 18.25 & 60.21 & 39.23 & 89.45 \\
			\bottomrule
		\end{tabular}
	\end{table*}

\begin{figure}[htp]
		\centering
		\sidesubfloat[][\label{fig:0p}]{
			\includegraphics[width=0.40\linewidth,
			trim=4cm 4cm 2cm 8.5cm,clip]{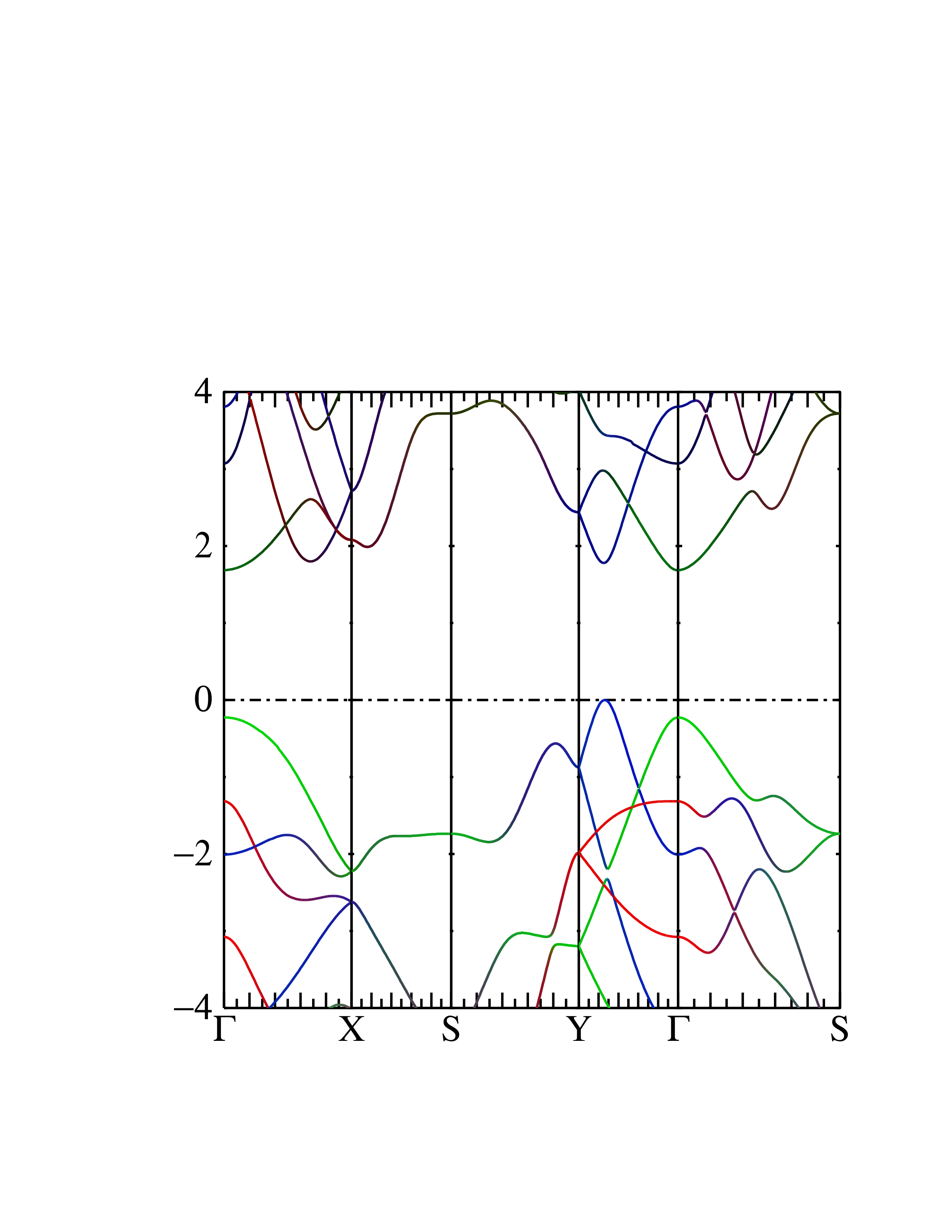}
		}
		\sidesubfloat[]{
			\includegraphics[width=0.40\linewidth,
			trim=4cm 4cm 2cm 8.5cm,clip]{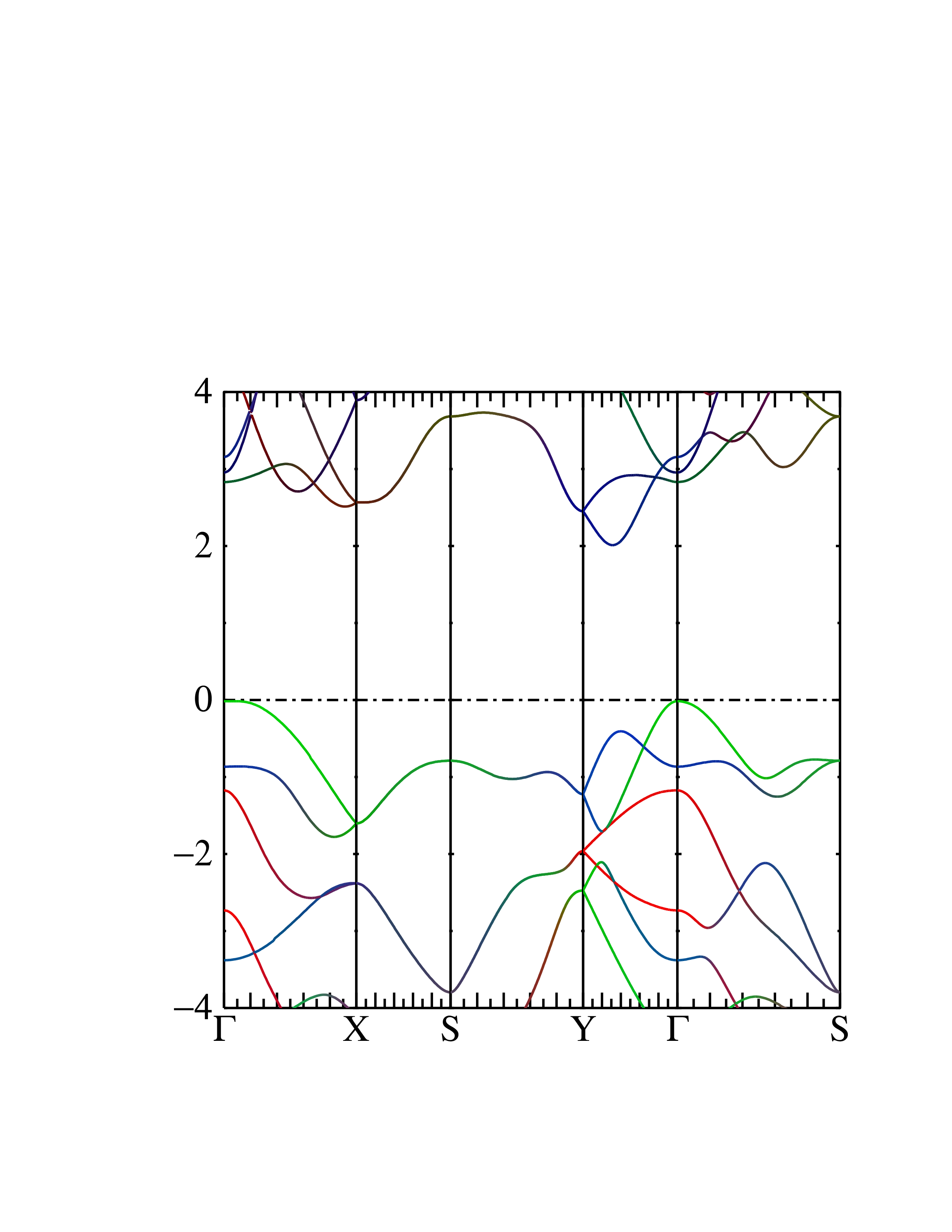}
		}\\
		\sidesubfloat[]{
			\includegraphics[width=0.40\linewidth,
			trim=4cm 4cm 2cm 8.5cm,clip]{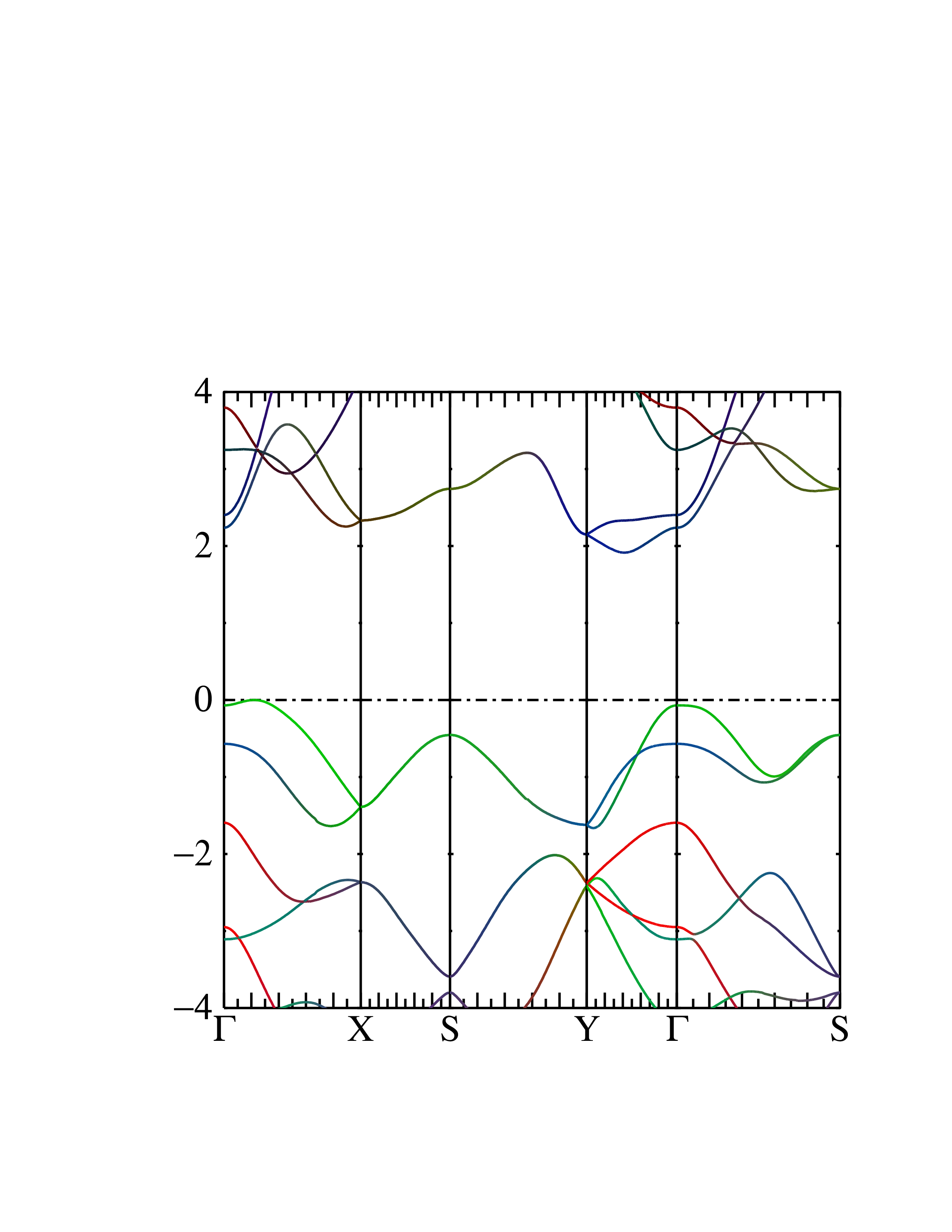}
		}
		\sidesubfloat[]{
			\includegraphics[width=0.40\linewidth,
			trim=4cm 4cm 2cm 8.5cm,clip]{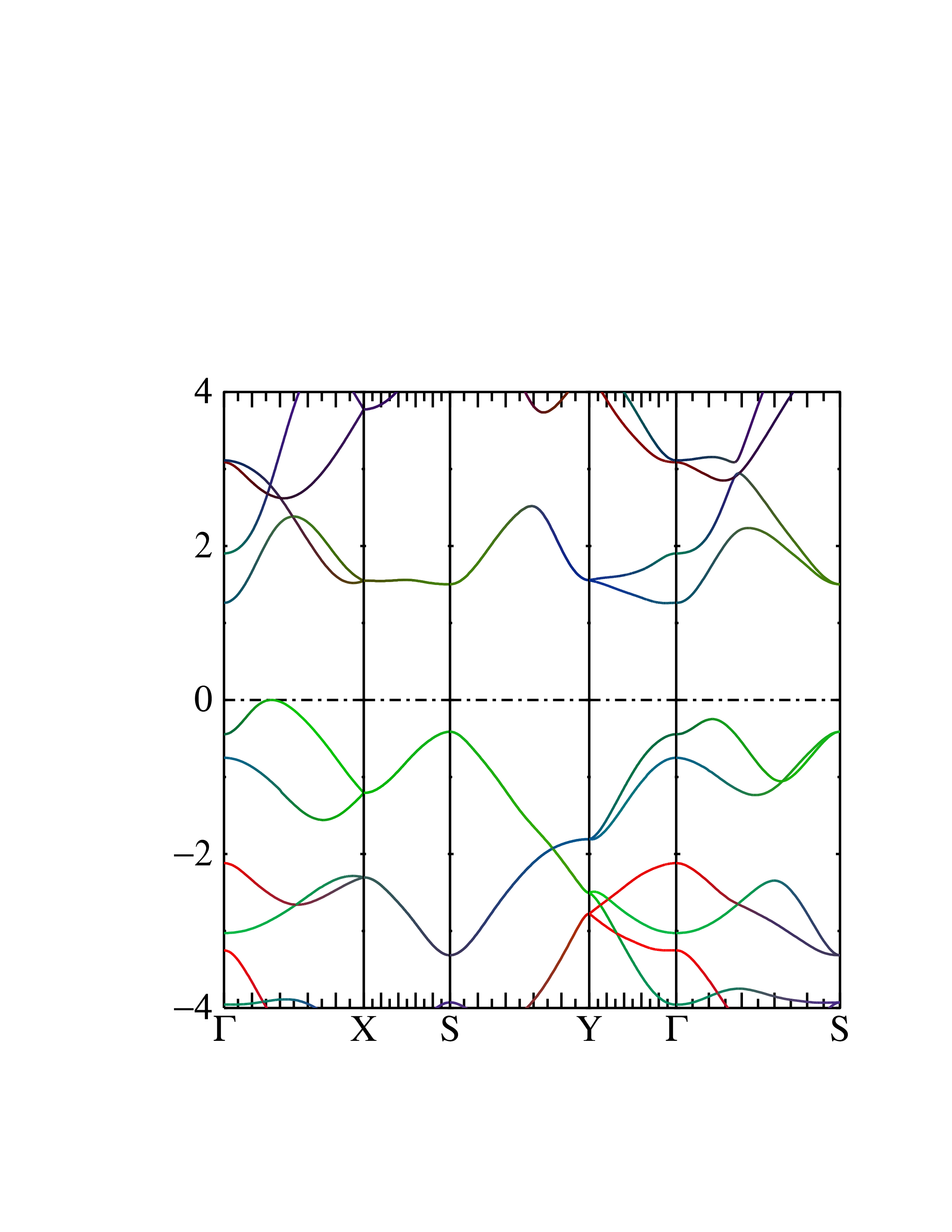}
		}\\
		\sidesubfloat[]{
			\includegraphics[width=0.40\linewidth,
			trim=4cm 4cm 2cm 8.5cm,clip]{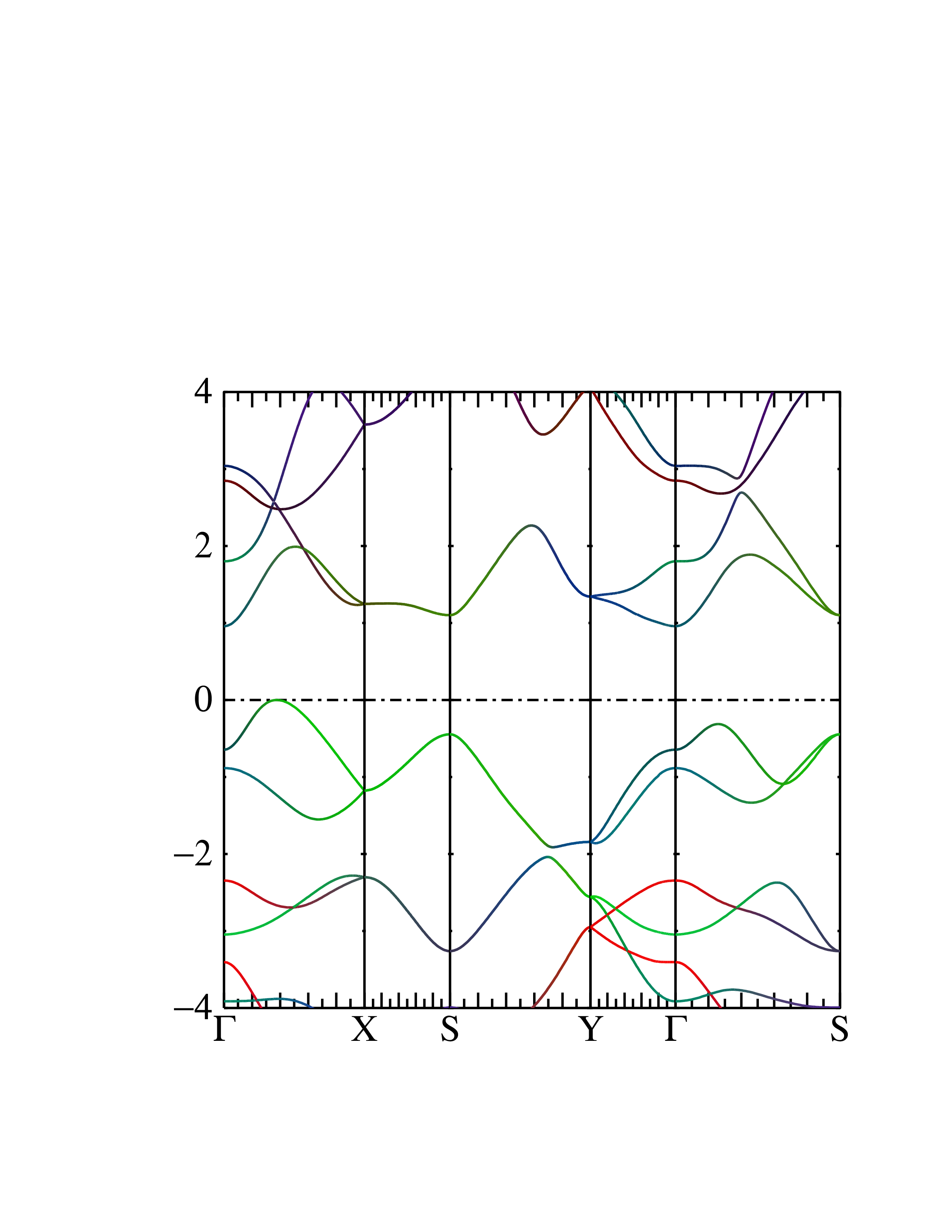}
		}
		\sidesubfloat[]{
			\includegraphics[width=0.40\linewidth,
			trim=4cm 4cm 2cm 8.5cm,clip]{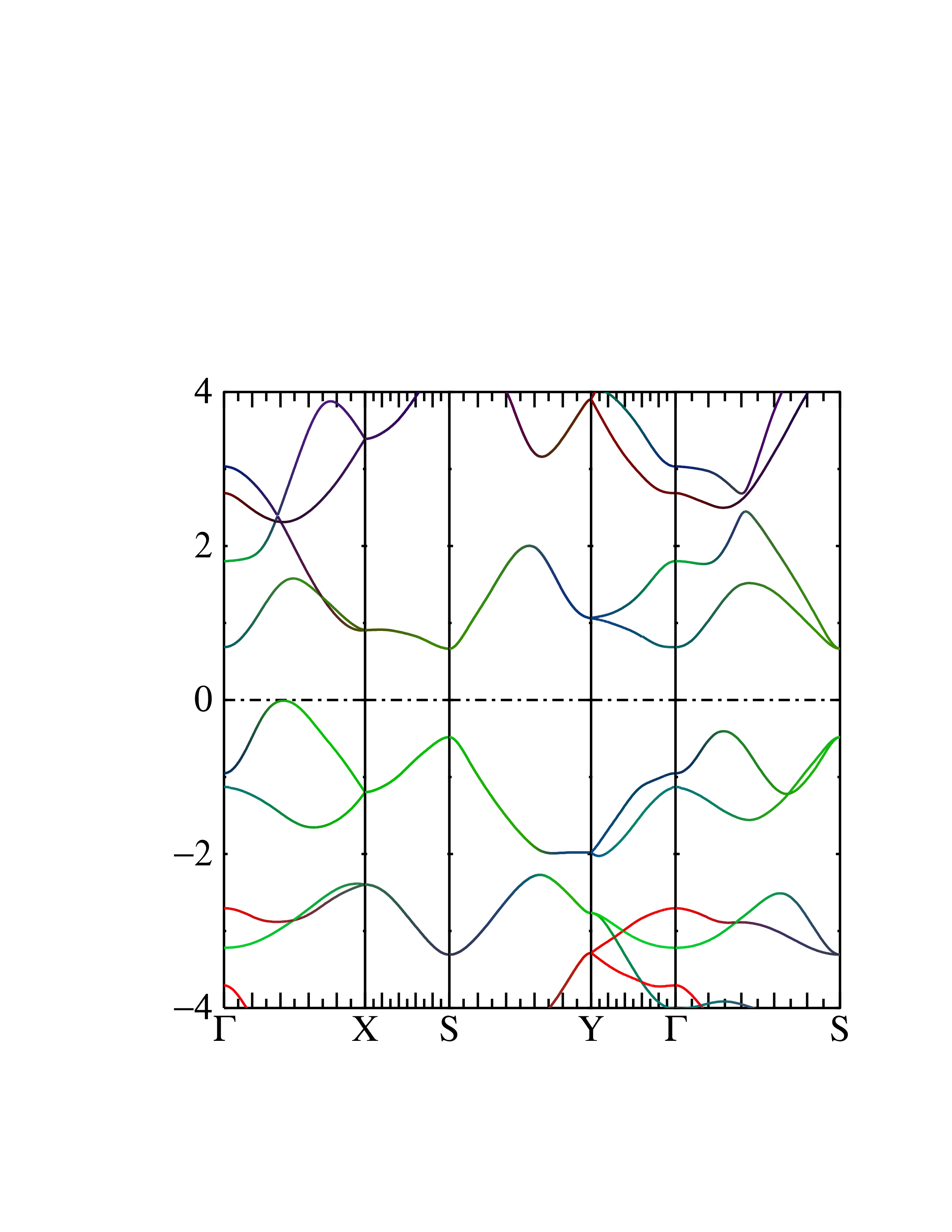}
		}
		\caption{Band structures at the QS$GW$ level for increasing biaxial strain, from
			(a) to (m). Orbital character is color-coded: green for $p_z$, blue for
			$p_y$, and red for $p_x$ states. The vertical axis shows energy in eV.
			Panels correspond to structures: (a)~0, (b)~8, (c)~16, (d)~23,
			(e)~25, and (f)~27. The $y$-axis shows energy (eV), and the $x$-axis follows the chosen high-symmetry path.
		}
		\label{fig:BandsEvolution1}
	\end{figure}

\begin{figure}[htp]
		\centering
		\setcounter{subfigure}{6}
		\sidesubfloat[]{
			\includegraphics[width=0.40\linewidth,
			trim=4cm 4cm 2cm 8.5cm,clip]{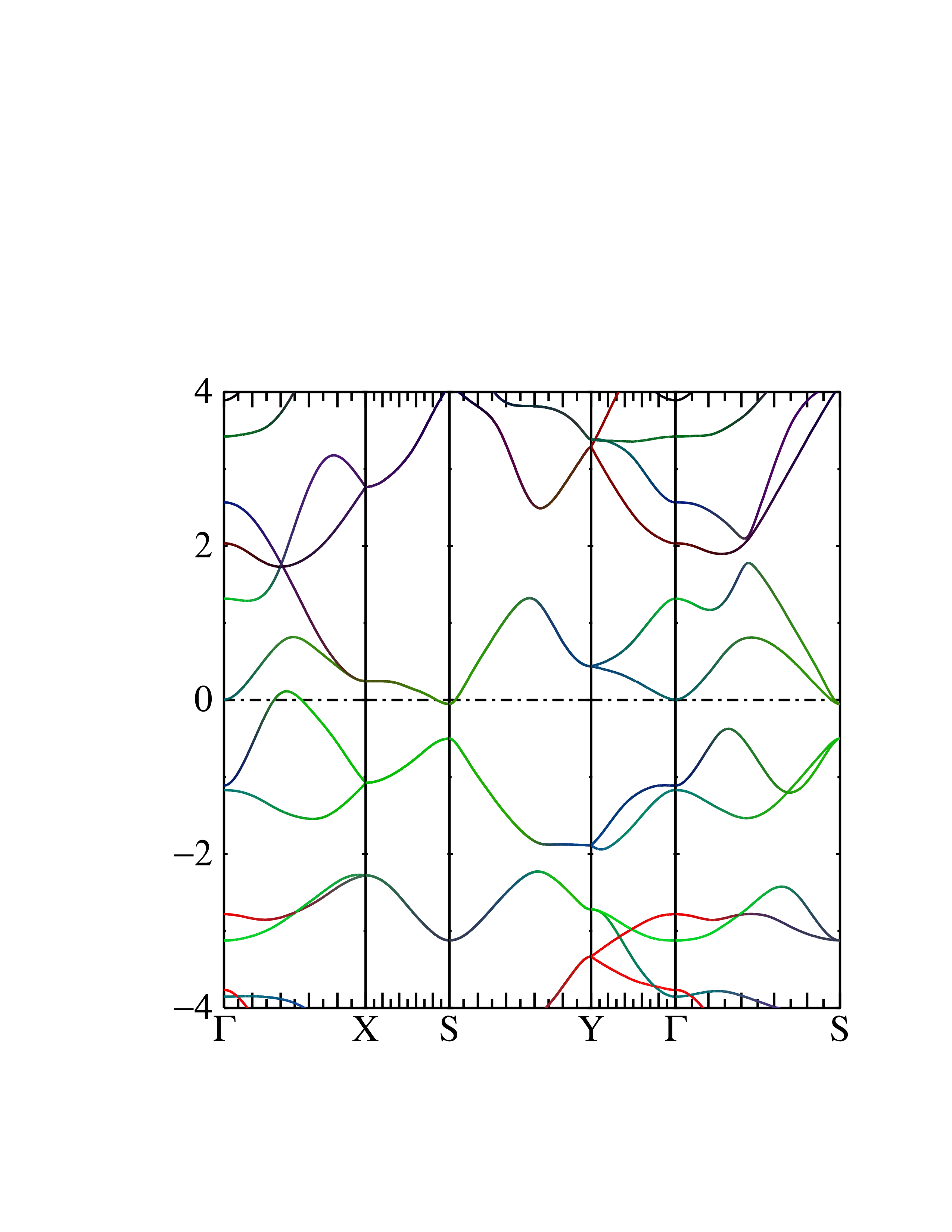}
		}
		\sidesubfloat[]{
			\includegraphics[width=0.39\linewidth,
			trim=4cm 4cm 2cm 8.5cm,clip]{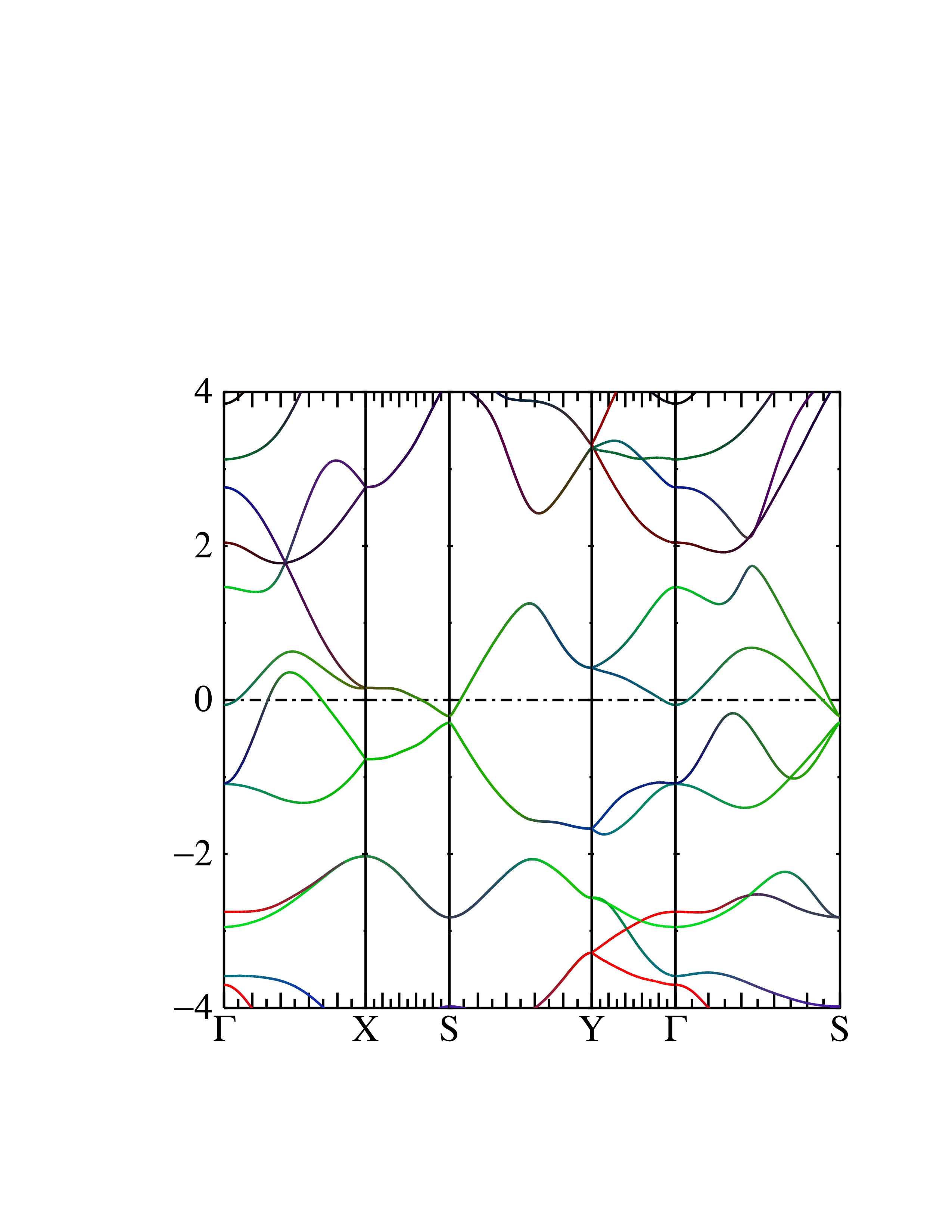}
		}\\
		\sidesubfloat[]{
			\includegraphics[width=0.40\linewidth,
			trim=4cm 4cm 2cm 8.5cm,clip]{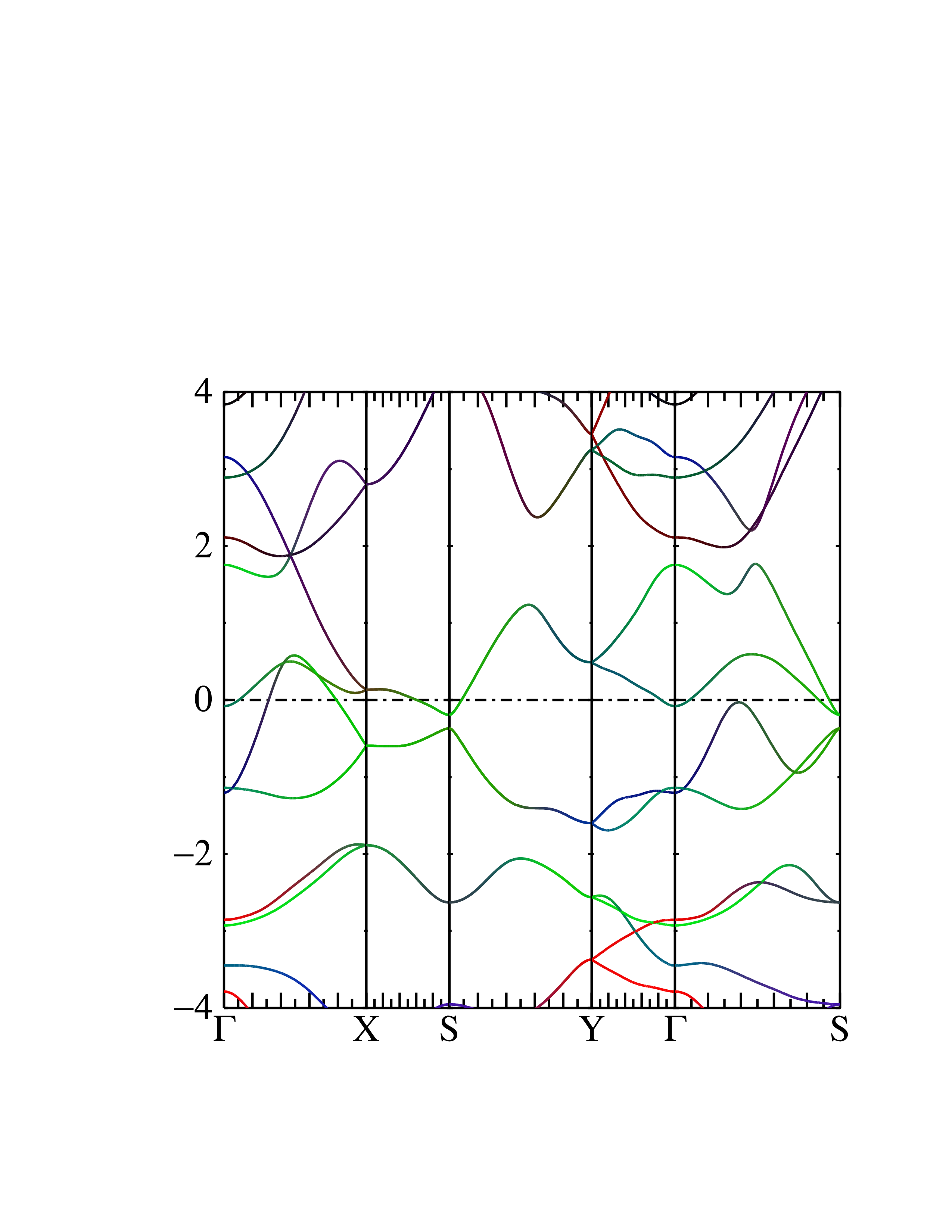}
		}
		\sidesubfloat[]{
			\includegraphics[width=0.40\linewidth,
			trim=4cm 4cm 2cm 8.5cm,clip]{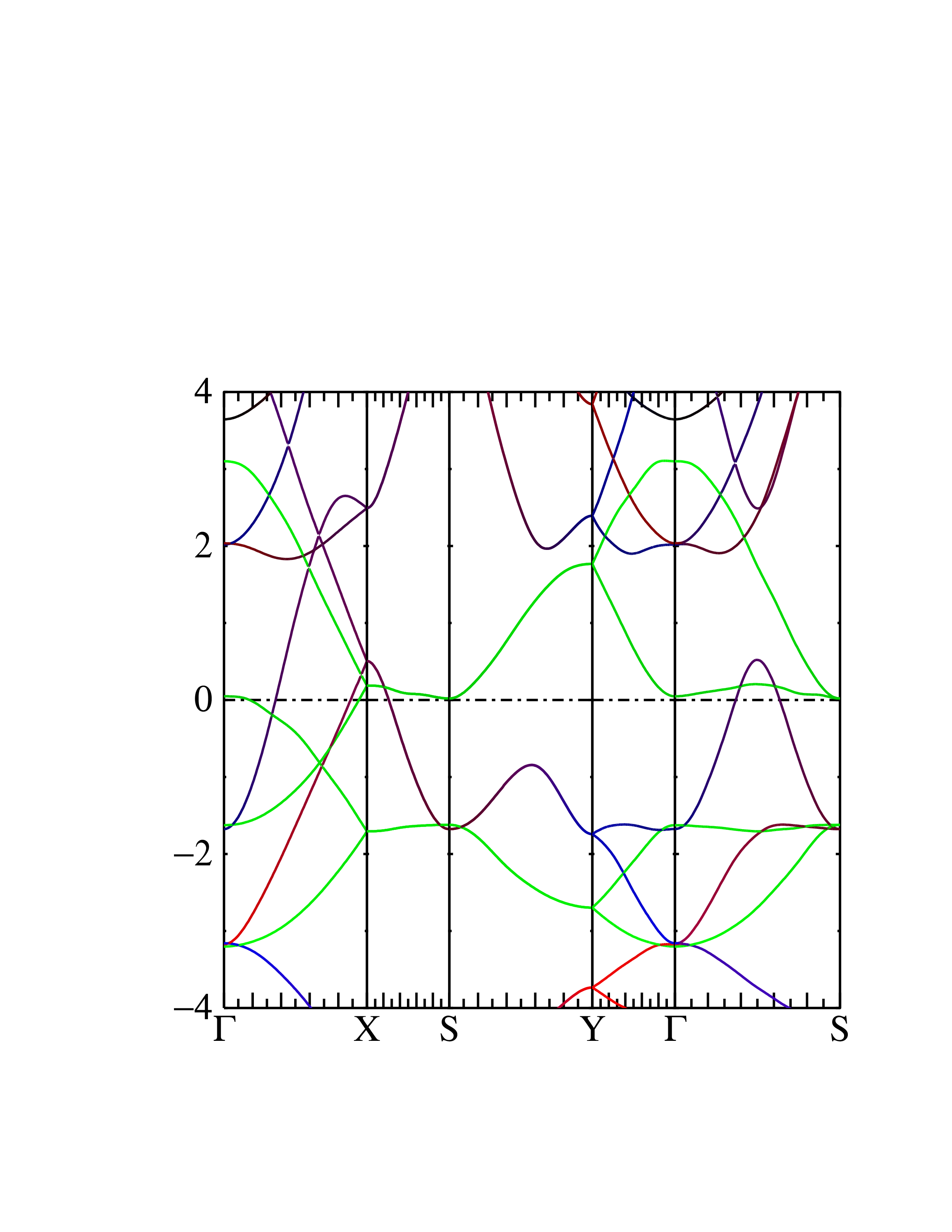}
		}
		\caption{Band structures at the QS$GW$ level for increasing biaxial
			strain, from (g) to (j). Orbital character is color-coded: green for
			$p_z$, blue for $p_y$, and red for $p_x$ states. The vertical axis shows
			energy in eV. Panels correspond to structures (g)~28, (h)~29, (i)~30,
			and (j)~32. The $y$-axis shows energy (eV), and the $x$-axis follows the chosen high-symmetry path.}
		\label{fig:BandsEvolution2}
	\end{figure}

	Next, we examine the evolution of the band structure along this
	relaxation path. Representative QS$GW$ band structures at selected
	strain steps are shown in Figures~\ref{fig:BandsEvolution1} and~\ref{fig:BandsEvolution2}.
	The orbital
	characters are shown using a color code: green for $p_z$, blue for $p_y$,
	and red for $p_x$.
	In Table~\ref{tab:biaxial-strain}, we list all sampled biaxial strain
	values. The structures highlighted in gray correspond to the specific
	steps shown in Figures~\ref{fig:BandsEvolution1} and~\ref{fig:BandsEvolution2}.
	
	First we start with Fig.~\ref{fig:BandsEvolution1}. In panel~(a), which
	corresponds to the equilibrium puckered phase, the valence-band maximum
	along $\Gamma$–Y has predominantly $p_y$ character and lies above the
	$p_z$-type local maximum at $\Gamma$. In panel~(b), these features have
	already switched order: the $p_z$-like valence and conduction bands at
	$\Gamma$ shift upward in energy, and the conduction-band minimum moves to
	a point along $\Gamma$–Y. In panel~(c), the top two valence bands detach
	from the lower manifold, and the $p_y$ and $p_z$ states begin to
	hybridize more strongly.
	In panel~(d), the conduction-band minimum has moved back to $\Gamma$,
	while the valence-band maximum shifts to a point along $\Gamma$–X. In
	intermediate steps, the changes are gradual and do not introduce
	qualitatively new features.
	
	By panel~(g), now shown in Fig.~\ref{fig:BandsEvolution2}, the lowest conduction band has
	dropped in energy at $\Gamma$, X, and S, and the gap has closed. The
	system becomes semimetallic, with a small hole pocket forming along
	$\Gamma$–X. In panel~(h), the gap between valence and conduction bands at S, which has gradually taken on a more $p_z$-like character, opens again (closed in a step prior, which is not shown here).
	
	It is also worth mentioning that,
	as the biaxial strain increases, the crossings of different bands changes; i.e. they are "unpinned". In particular, along the Y–$\Gamma$ path, the crossing between $p_y$
	and $p_z$ states, close to the Fermi energy gradually disappears as the $p_y$ bands hybridize increasingly with $p_z$, giving rise to a new crossing point
	at $\Gamma$. In contrast to $p_y$- and $p_z$-like bands, the
	$p_x$ character remains largely unaffected throughout the strain range,
	reflecting the anisotropic nature of the structural deformation.
	
\begin{figure}[htp]
		\centering
		\sidesubfloat[]{
			\includegraphics[width=.45\linewidth,
			trim=3cm 3cm 1cm 8cm,clip]{32p.png}
		}
		\sidesubfloat[]{
		\includegraphics[width=.4\linewidth]{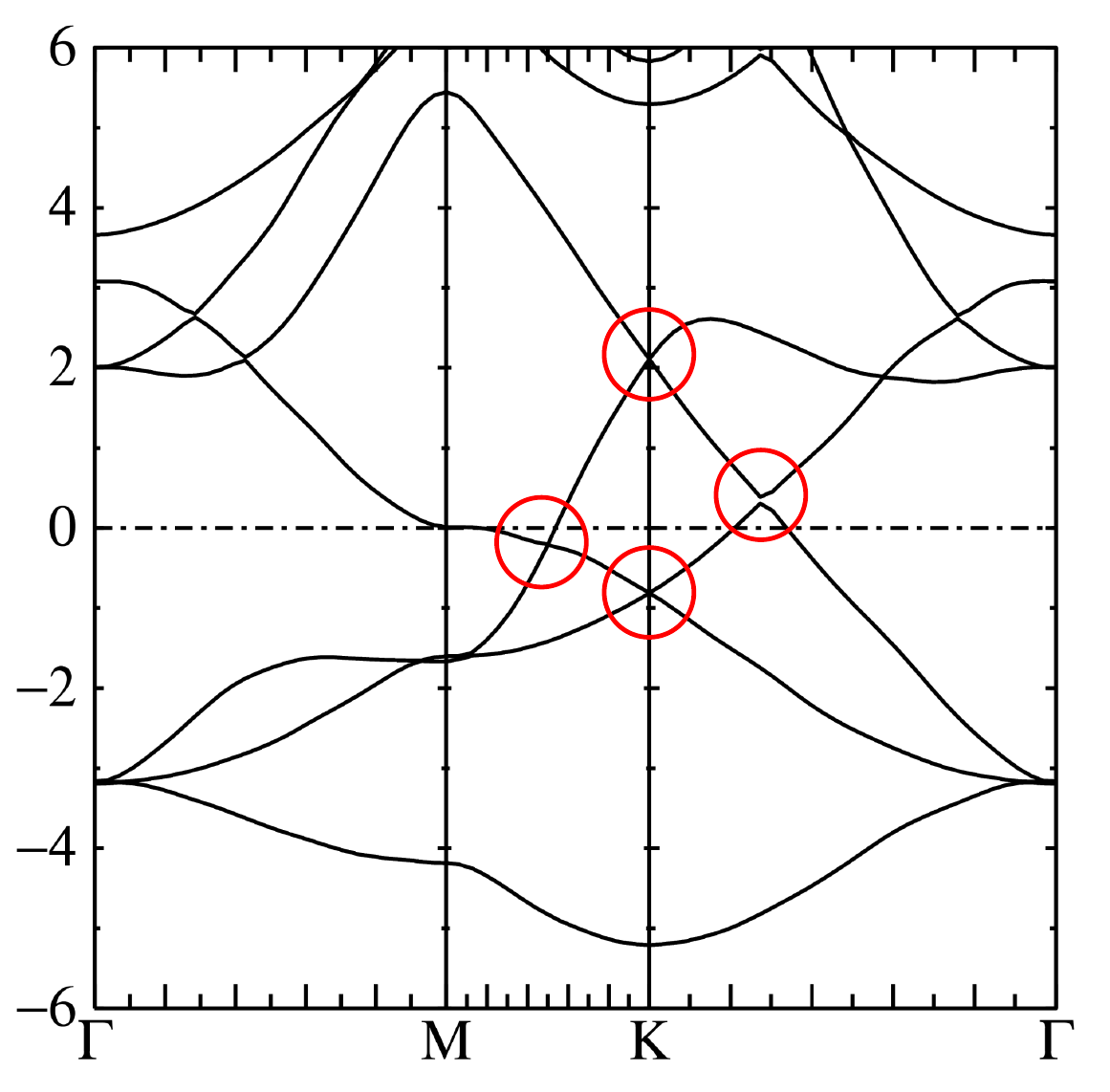}
		}\\
		\sidesubfloat[]{
			\includegraphics[width=.42\linewidth,
			trim=0cm 0cm 0cm 0cm,clip]{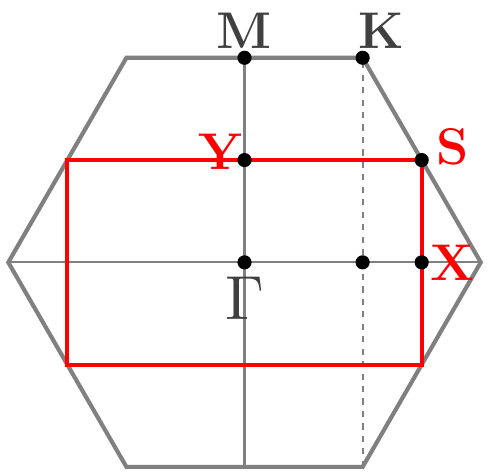}
		}
		\caption{(a) QS$GW$ band structure of the flat phase in the doubled
			orthorhombic unit cell.
			(b) QS$GW$ bands corresponding to panel (a), but plotted in the primitive
			hexagonal BZ. Several interesting crossings are indicated by red circles.
			(c) Brillouin-zone folding scheme showing
			how the orthorhombic BZ overlaps with the hexagonal BZ. In (a) and (b), the $y$-axis shows energy (eV), and the $x$-axis follows the chosen high-symmetry path.
			}
		\label{fig:Folding}
	\end{figure}

	In the final flat configuration, shown in panel (j), the band structure
	of the doubled orthorhombic cell can be related to that of the primitive
	hexagonal honeycomb phase through zone folding. This correspondence is
	illustrated in Fig.~\ref{fig:Folding}. As shown there, several of the
	interesting band crossings identified for antimonene in panel (c) can be
	mapped directly into the orthorhombic Brillouin zone. For example, the
	K point of the hexagonal BZ folds onto a point along the $\Gamma$–X
	direction of the orthorhombic BZ. Consequently, the $p_z$-like crossing
	at K (blue in panel (c)), located around $-1$~eV in panel (d), matches
	the corresponding crossing in panel (a) along $\Gamma$–X at approximately at the same $-1$~eV, where it also exhibits predominantly $p_z$ (green) character.
	
\begin{figure}
	\includegraphics[width=\columnwidth]{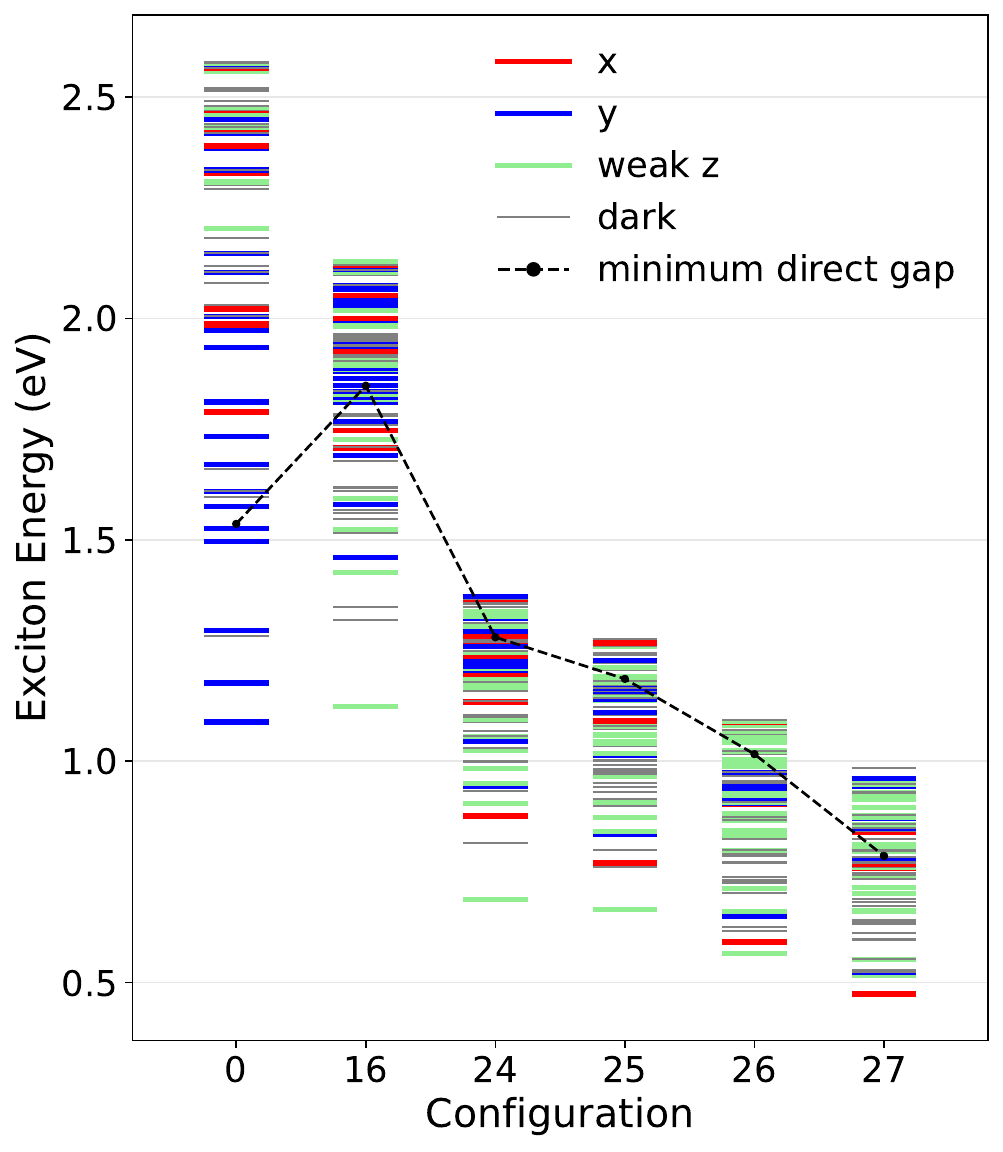}
		\caption{Excitonic energies for six different configurations, together with their classification as bright or dark and their polarization.}
		\label{fig:excitons_all}
	\end{figure}
	\begin{table}[h!]
		\centering
		\begin{tabular}{cccc}
			\toprule
			Config & LDA (eV) & QS$GW$ (eV) & QS$G\hat{W}$ (eV) \\
			\midrule
			0  & 0.7204 & 1.6761 & 1.4173 \\
			8  & 0.8094 & 2.0119 & 1.6459 \\
			16 & 0.6574 & 1.9148 & 1.4687 \\
			19 & 0.4525 & 1.7087 & -- \\
			23 & 0.1297 & 1.2381 & -- \\
			24 & 0.0528 & 1.0994 & 0.7499 \\
			25 & 0      & 0.9588 & 0.6304 \\
			26 & 0      & 0.8799 & 0.4830 \\
			27 & 0      & 0.6252 & 0.2998 \\
			28--32 & 0  & 0      & -- \\
			\bottomrule
		\end{tabular}
		\caption{Minimum band gaps as a function of configuration.}
		\label{tab:mingaps}
	\end{table}

	Next, we discuss the changes in optical properties and excitons under strain. A series of excitons occurs below the band gap. We classify these as $x$-polarized, $y$-polarized, weakly $z$-polarized, or dark depending on their oscillator strengths. If all oscillator strengths are lower than a threshold of $10^{-4}$ we call it dark. Fig. \ref{fig:excitons_all} shows the evolution of these exciton spectra as the structure changes referring to the configurations defined earlier. One can see the lowest excitonic state changes from $x$ polarization to weak $z$ to eventually $y$ polarization in this series of structures and these changes are related to the changes in band orbital nature near the gap  described earlier. We note that the minimum gap shown in Table \ref{tab:mingaps} closes beyond configuration 28 and BSE calculations for the resulting metallic systems are not meaningful. 
	Furthermore the excitons calculated here all correspond to direct optical transitions. Therefore we also need to examine what the minimum direct gap is for each of these configurations. This minimum direct gap is shown as the dashed black line in Fig.\ref{fig:excitons_all}. 
	
	\section{Conclusion}
	
	In this paper, we have characterized the electronic and optical
	properties of monolayer $\alpha$-arsenene and analyzed how the electronic properties evolve
	under strain on a biaxial strain path toward a flat honeycomb geometry. Using LDA,
	QS$GW$, and QS$G\hat{W}$ calculations, we showed that the relaxed
	puckered phase is a semiconductor with a direct gap in LDA that becomes
	indirect once quasiparticle corrections are included. The puckered structure has doubly degenerate bands
	along the Brillouin-zone edge X-S-Y in the absence of SOC, and
	SOC lifts these degeneracies everywhere except at the X and Y points due to the non-symmorphic
	symmetry of the structure, leading to
	symmetry-protected Dirac points. Orbital-resolved band
	structures revealed the separation of As-$s$ states from the $p$-derived
	valence and conduction manifolds and a $k$-dependent $p$–$d$
	hybridization that is enhanced at the QS$GW$ level, particularly near S.
	This orbital analysis provided a way for tracking band
	reordering and hybridization along the strain path, which was discussed in the second part of this paper.
	
	The optical response of $\alpha$-arsenene, obtained from BSE
	calculations, was found to be strongly anisotropic, reflecting the anisotropy of the puckered structure. For light polarized along $x$ and $y$, the
	independent-particle spectra are significantly modified once
	electron–hole interactions are included, with a clear red shift of the
	absorption onset. The first bright exciton appears at different energies
	for the two in-plane polarizations and carries a noticeably weaker
	oscillator strength along $x$. The corresponding BSE eigenvectors modulus projected on the bands show
	that these excitons are built mainly from transitions between the
	highest valence and lowest conduction bands, with distinct $\mathbf{k}$-space distribution for the brighter $y$-polarization excitons compared to the darker x-polarization excitons.
	We also noted the presence of dark excitons
	and that, while the lowest exciton is bright in the relaxed puckered
	structure, it becomes dark when strain is applied.
	
	By performing structural relaxations under uniaxial and
	biaxial strain, we found that only biaxial in-plane strain can fully
	flatten the puckered layer into a honeycomb-like configuration.
	Along this biaxial path, the in-plane and out-of-plane bond lengths
	approach each other, the puckering height collapses, and the bond angles
	tend toward $120^\circ$, signaling the formation of a planar network.
	The accompanying QS$GW$ band structures exhibit a sequence of changes:
	the location and character of the band edges shift, the gap closes and
	reopens, and the crossings between $p_y$- and $p_z$-like bands become
	“unpinned” and migrate in the Brillouin zone. In the final flat phase,
	the band structure of the doubled orthorhombic cell can be related to
	the primitive honeycomb Brillouin zone by zone folding, and several Dirac crossings known from related group-V monolayers can be identified in
	this representation. Finally, we traced the evolution of the exciton series and its polarization character as function of biaxial strain. 
	Together, these results show that $\alpha$-arsenene
	is a highly strain-tunable monolayer, in which
	non-symmorphic symmetry, spin–orbit coupling, and excitonic effects all
	play an important role in shaping the electronic and optical
	properties.
	\acknowledgments{This work was supported by the US Department of Energy, Basic Energy Sciences under grant number DE-SC000893 and made use of the High Performance Computing Resource in the Core Facility for Advanced Research Computing at Case Western Reserve University.}

{\bf DATA  AVAILABILITY}
The data that support the findings of this article will be made publicly available.
	
	
	\bibliography{combined}
	\bibliographystyle{apsrev4-1}

	\end{document}